\documentclass[reprint,amsmath,amsfonts,amssymb,aps,prx,preprintnumbers,superscriptaddress]{revtex4-2}

\usepackage[utf8]{inputenc}
\usepackage[T1]{fontenc}
\usepackage{lmodern}
\usepackage{graphicx}
\usepackage{dcolumn}
\usepackage{bm}
\usepackage{textcomp}
\usepackage{ulem}
\usepackage{ifpdf}
\usepackage[squaren,Gray]{SIunits}
\usepackage{color}
\definecolor{red}{rgb}{1,0,0}
\definecolor{blue}{rgb}{0,0,1}
\definecolor{darkred}{rgb}{0.6,0,0}
\definecolor{darkblue}{rgb}{0,0,0.6}
\definecolor{darkgreen}{rgb}{0,0.5,0}
\definecolor{grey}{rgb}{0.7,0.5,0.8}

\ifpdf
\usepackage{epstopdf}
\usepackage[pdftex,unicode,pdfstartview={FitH},pdfborder={0 0 0}]{hyperref}
\usepackage{hypcap}
\else
\usepackage[hypertex]{hyperref}
\fi
\hypersetup{
    bookmarksnumbered = true,
    colorlinks = true, linkcolor = darkblue,
    citecolor = darkblue, filecolor = darkblue,
    menucolor = darkblue, urlcolor = darkblue
}

\newcolumntype{R}{>{$\displaystyle}r<{$}}
\newcolumntype{C}{>{$\displaystyle}c<{$}}

\begin{document}

\title{Metasurface of strongly coupled excitons \\
and nanoplasmonic arrays}

\author{Farsane Tabataba-Vakili}
\affiliation{Fakult\"at f\"ur Physik, Munich Quantum Center, and
  Center for NanoScience (CeNS), Ludwig-Maximilians-Universit\"at
  M\"unchen, Geschwister-Scholl-Platz 1, 80539 M\"unchen, Germany}
\affiliation{Munich Center for Quantum Science and Technology (MCQST),
  Schellingtra\ss{}e 4, 80799 M\"unchen, Germany}
  
\author{Lukas Krelle}
\affiliation{Fakult\"at f\"ur Physik, Munich Quantum Center, and
  Center for NanoScience (CeNS), Ludwig-Maximilians-Universit\"at
  M\"unchen, Geschwister-Scholl-Platz 1, 80539 M\"unchen, Germany}
\affiliation{Now at: Institute for Condensed Matter Physics, Technische Universität Darmstadt, Hochschulstraße 6, 64289 Darmstadt, Germany }

\author{Lukas Husel}
\affiliation{Fakult\"at f\"ur Physik, Munich Quantum Center, and
  Center for NanoScience (CeNS), Ludwig-Maximilians-Universit\"at
  M\"unchen, Geschwister-Scholl-Platz 1, 80539 M\"unchen, Germany}  

\author{Huy P. G. Nguyen}
\affiliation{Fakult\"at f\"ur Physik, Munich Quantum Center, and
  Center for NanoScience (CeNS), Ludwig-Maximilians-Universit\"at
  M\"unchen, Geschwister-Scholl-Platz 1, 80539 M\"unchen, Germany}

\author{Zhijie Li}
\affiliation{Fakult\"at f\"ur Physik, Munich Quantum Center, and
  Center for NanoScience (CeNS), Ludwig-Maximilians-Universit\"at
  M\"unchen, Geschwister-Scholl-Platz 1, 80539 M\"unchen, Germany}
  
\author{Ismail Bilgin}
\affiliation{Fakult\"at f\"ur Physik, Munich Quantum Center, and
  Center for NanoScience (CeNS), Ludwig-Maximilians-Universit\"at
  M\"unchen, Geschwister-Scholl-Platz 1, 80539 M\"unchen, Germany}
  
\author{Kenji Watanabe}
\affiliation{Research Center for Functional Materials, National Institute for Materials Science, 1-1 Namiki, Tsukuba 305-0044, Japan}

\author{Takashi Taniguchi}
\affiliation{International Center for Materials Nanoarchitectonics, 
National Institute for Materials Science, 1-1 Namiki, Tsukuba 305-0044, Japan}

\author{Alexander H\"ogele}
\affiliation{Fakult\"at f\"ur Physik, Munich Quantum Center, and
  Center for NanoScience (CeNS), Ludwig-Maximilians-Universit\"at
  M\"unchen, Geschwister-Scholl-Platz 1, 80539 M\"unchen, Germany}
\affiliation{Munich Center for Quantum Science and Technology (MCQST), Schellingtra\ss{}e 4, 80799 M\"unchen, Germany}

\begin{abstract}
Metasurfaces allow to manipulate light at the nanoscale. Integrating metasurfaces with transition metal dichalcogenide monolayers provides additional functionality to ultrathin optics, including tunable optical properties with enhanced light-matter interactions. In this work, we demonstrate the realization of a polaritonic metasurface utilizing the sizable light-matter coupling of excitons in monolayer WSe$_2$ and the collective lattice resonances of nanoplasmonic gold arrays. To this end, we developed a novel fabrication method to integrate gold nanodisk arrays in hexagonal boron nitride and thus simultaneously ensure spectrally narrow exciton transitions and their immediate proximity to the near-field of array surface lattice resonances. In the regime of strong light-matter coupling, the resulting van der Waals metasurface exhibits all key characteristics of lattice polaritons, with a directional and linearly-polarized far-field emission profile dictated by the underlying nanoplasmonic lattice. Our work can be straightforwardly adapted to other lattice geometries, establishing structured van der Waals metasurfaces as means to engineer polaritonic lattices. 
\end{abstract}

\maketitle

Ultrathin optical components known as metasurfaces have the potential to strongly impact modern optics by offering new functionality and unprecedented compactness. Comprised of periodic arrays of sub-wavelength nanostructures, metasurfaces modulate the amplitude, phase, or polarization of incident light \cite{yu2014flat, meinzer2014plasmonic, chen2016review}.  Plasmonic metasurfaces consisting of metallic nanostructures provide strong field enhancement and have been used to introduce abrupt phase changes \cite{yu2011light}, create holograms \cite{huang2013three}, and realize ultrathin lenses \cite{aieta2012aberration}. While the localized surface plasmon resonances of individual nanoparticles suffer from dephasing and dissipative losses, periodic arrays of such nanoparticles with lattice constants in the order of the wavelength allow to couple the plasmon resonance to the diffractive orders of the lattice, giving rise to collective surface lattice resonances (SLRs) with quality factors exceeding $2000$ at telecom wavelengths \cite{deng2020ultranarrow} and geometry-specific, angle-dependent dispersions \cite{cherqui2019plasmonic,guo2017geometry}. 

Van der Waals semiconductors are ideal building blocks for adding new functionality to metasurfaces due to their rich optical properties \cite{meng2023functionalizing}, intrinsic two-dimensionality and ease of integration by viscoelastic stamping \cite{pizzocchero2016hot}. Monolayers of transition metal dichalcogenides (TMDs) are particularly appealing due to direct bandgaps \cite{splendiani2010emerging,mak2010atomically}, large exciton binding energies \cite{he2014tightly,chernikov2014exciton}, high oscillator strengths \cite{zhang2014absorption,li2014measurement}, and valley-selective chiral optical transitions \cite{xiao2012coupled,mak2012control,wang2018colloquium}, with recent demonstrations of resonance tuning of atomically thin lenses \cite{van2020exciton}, separation of valley excitons \cite{sun2019separation}, and strong light-matter coupling with dielectric \cite{dufferwiel2015exciton,liu2015strong,chen2020metasurface} and plasmonic cavities \cite{liu2016strong,wang2019limits,liu2019observation}. However, integrating TMDs with plasmonic nanostructures without compromising the optical quality of pristine TMD monolayers is challenging, as they exhibit a strong sensitivity to strain \cite{khatibi2018impact} and dielectric disorder \cite{raja2019dielectric} due to their two-dimensional nature. 

\begin{figure*}[t!]
\includegraphics[scale=1]{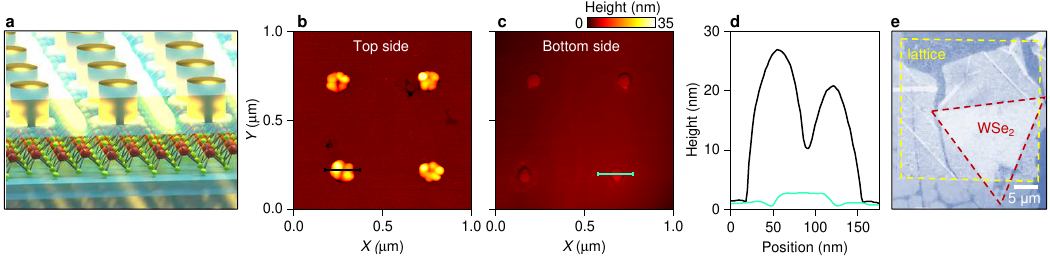}
\caption{ \textbf{Plasmon-exciton-polariton van der Waals metasurface.} \textbf{a}, Illustration of the metasurface and the near-field of the plasmonic surface lattice resonance. The sample consists of a TMD monolayer sandwiched between two hBN flakes, with a gold nanodisk lattice embedded in the top hBN layer. \textbf{b} and \textbf{c}, Atomic force micrographs of the top and bottom sides of a gold nanodisk array imprinted in hBN. The bottom side was measured with the hBN lattice flake on a PDMS/PCL stamp. \textbf{d}, Height profiles of individual gold nanodisks indicated by the black and green lines in \textbf{b} and \textbf{c}. \textbf{e}, False-color optical micrograph of the sample, with dashed lines indicating the lattice and the WSe$_2$ monolayer.} \label{fig1}
\end{figure*}

The prime strategy for reducing the detrimental effects of strain and environmental disorder on the optical properties of TMD monolayers is provided by encapsulation between layers of hexagonal boron nitride (hBN) that ensure an atomically flat and clean dielectric environment \cite{cadiz2017excitonic}. However, an hBN spacer between the TMD and the plasmonic nanoparticles would prevent immediate proximity to the near-field and thus reduce the coupling strength significantly. To date, this caveat has not been resolved in coupled exciton-plasmon systems, realizing either direct contact between plasmonic nanostructures and TMD monolayers void of hBN encapsulation \cite{wang2016coherent,liu2016strong,wang2019limits,liu2019observation,sun2021strong, guo2023hybrid} and thus subject to compromised optical quality, or using hBN-encapsulated monolayers on plasmonic nanostructures \cite{klein20192d, dibos2019electrically, vadia2023magneto} without immediate access to the near-field. 

In this work, we demonstrate strong coupling in a TMD-based plasmon-exciton-polariton metasurface utilizing a novel fabrication method. The approach we developed allows to integrate a gold nanodisk array directly into an hBN layer, and thus realize a van der Waals metasurface with immediate proximity between the near-field of the nanoplasmonic array and the TMD monolayer preserving the benefits of hBN-encapsulation. Both aspects of the integrated system ensure strong coupling between the TMD monolayer exciton and the SLR of the plasmonic nanodisc array, giving rise to plasmon-exciton-polaritons with large Rabi splitting and effective polariton dispersion conditioned by the geometry of the plamonic lattice. Consistently, the resulting emission profile of the lower polariton branch is strongly modified as compared to the emission characteristics of uncoupled monolayer excitons, exhibiting narrow-angle directional light emission with high degree of linear polarization, as dictated by the SLRs of a square lattice.

Our van der Waals metasurface, illustrated in Fig.~\ref{fig1}a, contains a WSe$_2$ monolayer synthesized by chemical vapor deposition \cite{li2022stacking} and high quality hBN-layers on both sides. In the top hBN layer, we incorporated a plasmonic gold nanodisk array by dry-etching air-holes into hBN and filling them with gold (details in Methods). Atomic force micrographs of the top and bottom sides of the hBN layer with the imprinted gold lattice are shown in Fig.~\ref{fig1}b and c, and the linecuts of the respective height profiles in Fig.~\ref{fig1}d. Clearly, the bottom side with a maximum change in height below $1$~nm is significantly smoother than the top side, where individual gold disks reach out above the hBN-surface by as much as $\sim 25$~nm. An additional key advantage of the bottom surface for further integration with the TMD monolayer is the fact that it was not directly exposed to any processing. Placing the bottom side on top of the TMD monolayer thus results in a flat and clean interface with immediate proximity to the near-field of the plasmonic SLRs. Moreover, the robustness of the lattice and adherent TMD allows consecutive pickup of the coupled system from the substrate via the hot-pickup technique \cite{pizzocchero2016hot}, as confirmed by the optical micrograph of the van der Waals heterostructure shown in Fig.~\ref{fig1}e. Subsequently, the stack was deposited on the terminating bottom hBN-layer on SiO$_2$/Si substrate.  

\begin{figure*}[t!]
\includegraphics[scale=1]{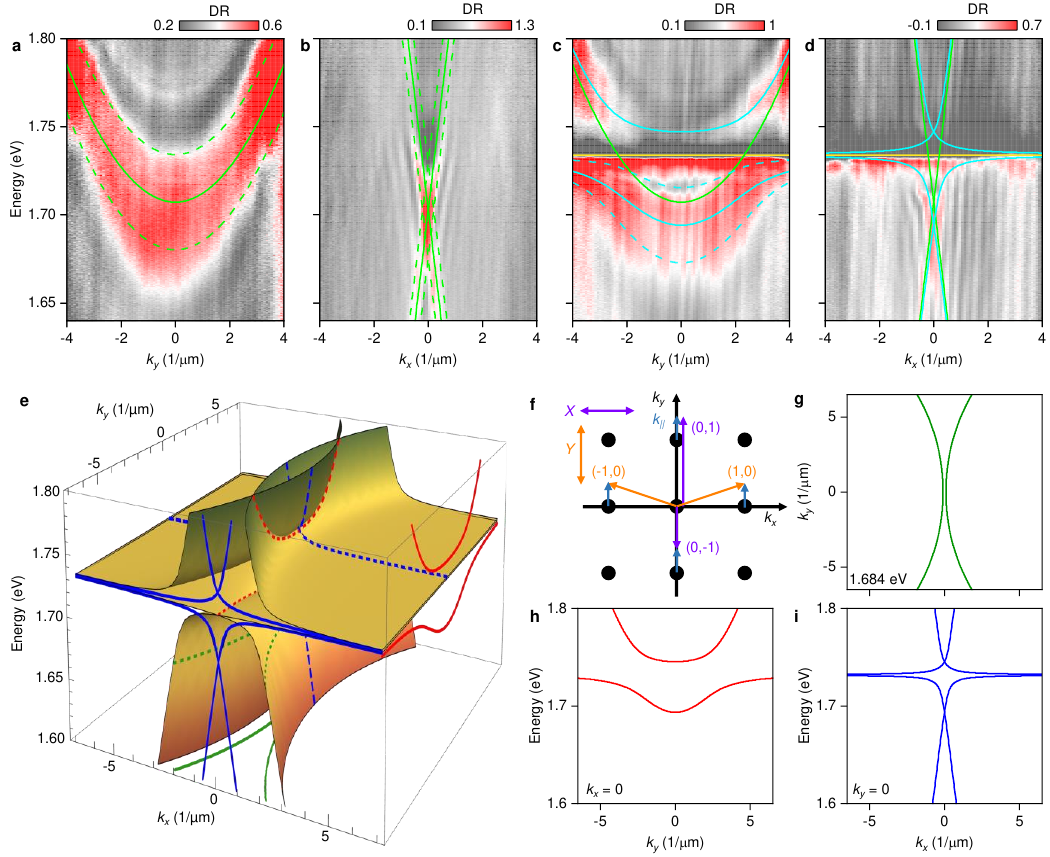}
\caption{ \textbf{Momentum-resolved lattice and polariton dispersions.}  \textbf{a} and \textbf{b},  Hyperbolic and linear  dispersions of the $(\pm 1,0)$ SLRs of the uncoupled lattice in differential reflectance (DR) with linear excitation along $Y$ and detection along $k_y$ and $k_x$, respectively. Solid (dashed) lines indicate the fitted dispersions (linewidth of $\gamma_\text{SLR}=27$~meV). \textbf{c} and \textbf{d}, DR polariton dispersions with linear excitation along $Y$ and detection along $k_y$ and $k_x$, respectively. The SLRs (green), bare exciton energy (yellow), and fitted polariton branches (cyan) are overlaid. Dashed cyan lines in \textbf{c} indicate the lower polariton linewidth $\gamma_{\text{LP}}$.  The light-matter coupling strength $g$ was determined as $25$~meV from the best fit. \textbf{e}, Three-dimensional plot of the plasmon-exciton-polariton energy dispersion of the $Y$-polarized SLRs branches with ($\pm1,0$) diffractive orders, according to Eq.~\eqref{ham}. \textbf{f}, Reciprocal lattice of a square lattice with in-plane $k$-vector $k_\parallel$ along $k_y$. The four lowest diffractive orders $(\pm 1,0)$ and $(0,\pm 1)$, as well as linear polarization along $X$ and $Y$ are indicated. \textbf{g -- i}, Two-dimensional projections of the dispersion in \textbf{e} for $E= 1.684$~eV (\textbf{g}), $k_x=0$ (\textbf{h}), and $k_y=0$ (\textbf{i}), with panels \textbf{h} and \textbf{i} corresponding to \textbf{c} and \textbf{d}, respectively.} \label{fig2}
\end{figure*}

First, we determined the dispersion of the bare gold lattice in hBN using differential reflectance (DR) spectroscopy with data in Figs.~\ref{fig2}a and b. The geometric model of the diffractive orders of the SLRs \cite{guo2017geometry}, illustrated in Fig.~\ref{fig2}f for momentum along the $k_y$ axis, identifies the four lowest diffractive orders $(\pm 1, 0)$ and $(0, \pm 1)$ of the SLRs. For the square lattice, linearly polarized excitations couple either to the $(\pm 1, 0)$ or $(0, \pm 1)$ diffractive orders, which show degenerate hyperbolic (Fig.~\ref{fig2}a) or non-degenerate linear (Fig.~\ref{fig2}b) energy dispersions depending on the combination of linear polarization and $k$-space detection axes. With the $k$-vector of the SLRs given by $k_{\text{SLR}} = k_{\text{DO}}+k_{\parallel}$, the energy dispersion is given by \cite{guo2017geometry}:
\begin{equation}\label{energy}
E_{\text{SLR}}(k_x,k_y)= \frac{\hbar c}{n_{\text{eff}}} \sqrt{\left( \frac{2n \pi}{a}+ k_x \right)^2+ \left( \frac{2m \pi}{a} + k_y \right)^2},
\end{equation}
with the lattice constant $a$, diffractive orders $(n,m)$, and effective refractive index $n_{\text{eff}}$. In the long-wavelength limit at $k=0$, $\lambda_\text{SLR}\approx n_\text{eff}a$. By fitting Eq.~\eqref{energy} to the dispersions measured at different positions on the uncoupled lattice and averaging over an area of $\sim 400~\mu$m$^2$, we determined $n_{\text{eff}} = 1.517\pm0.007$ and $E_\text{SLR}=1.70\pm0.01$~eV at $k=0$. The best-fit procedure also yields the half-width at half-maximum linewidth of the inhomogenously broadened Gaussian SLR in phase-corrected DR (obtained using the Kramers-Kronig relation) as $\gamma_\text{SLR}= 27$~meV, much narrower than for localized surface plasmons at the same energy \cite{vadia2023magneto} and corresponding to a quality factor of $32$. 

\begin{figure*}[t!]
\includegraphics[scale=1]{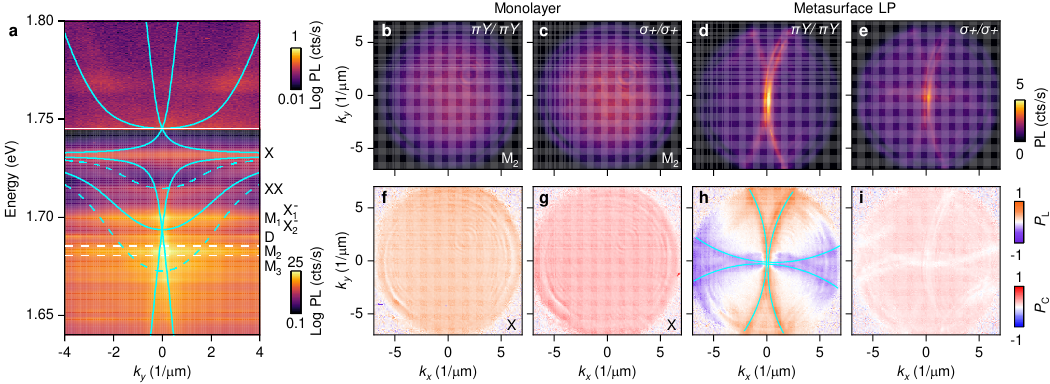}
\caption{\textbf{Momentum-resolved photoluminescence.} \textbf{a}, Momentum-dispersed PL from the metasurface region, recorded with linear excitation along $Y$ and shown together with the hyperbolic dispersions of the upper and lower polariton branches $E_{\text{LP,UP}}$ (solid cyan lines) and the inhomogeneous broadening $\gamma_\text{LP}$ of the lower polariton (dashed cyan lines). Note the different intensity scales in the top and bottom parts for better visualization of the upper polariton branch PL. \textbf{b} and \textbf{c}, Momentum-space PL images of the phonon sideband M$_2$ on a bare WSe$_2$ monolayer with Gaussian emission profile in both linear and circular polarization bases. The excitation was linear along $Y$ ($\pi Y$) or circular ($\sigma +$) and the detection was co-polarized, as indicated in each panel. \textbf{d} and \textbf{e}, Same but for the the metasurface in the energy range delimited by white dashed lines in \textbf{a}. \textbf{f} and \textbf{g}, Degrees of linear and circular polarization, $P_\text{L}$ and $P_\text{C}$, of the bright exciton X in monolayer WSe$_2$. \textbf{h} and \textbf{i}, Same but for the lower polariton branch of the metasurface, with dispersions of the four lowest diffractive orders shown by solid lines in \textbf{h}.}
\label{fig3}
\end{figure*}

In sample regions of the van der Waals metasurface, strong coupling between the plasmonic SLRs and the bright monolayer exciton X gives rise to exciton-polaritons, with the energy dispersion of the upper (E$_{\text{UP}}$) and lower (E$_{\text{LP}}$) polariton branches given by (see Methods for details):
\begin{equation}\label{ham}
    \begin{split}
    E_{\text{UP,LP}}(k_x,k_y) = \frac{E_{\text{SLR}} + E_{\text{X}} - i(\gamma_{\text{X}}-\gamma_{\text{SLR}})}{2} \\
    \pm \sqrt{g^{2} - \frac{1}{4}[E_{\text{X}} - E_{\text{SLR}} - i(\gamma_{\text{X}} - \gamma_{\text{SLR}})]^{2}},
    \end{split}
\end{equation}
where E$_{\text{X}}$ and $\gamma_{\text{X}}$ are the exciton energy and half-width at half-maximum linewidth, respectively, and $g$ denotes the light-matter coupling strength. Representative experimental dispersions of the coupled system, recorded in DR spectroscopy, are shown in Figs.~\ref{fig2}c and d.

By fitting Eq.~\eqref{ham} to the polariton dispersions recorded on different positions and averaging over an area of $\sim 300~\mu$m$^2$ of the metasurface, with the exciton linewidth $\gamma_{\text{X}}=3.5$~meV determined from Lorentzian fits to the phase-corrected DR on a monolayer region, we found $g=25\pm 2$~meV for the light-matter coupling strength, $n_\text{eff}=1.513 \pm 0.005$ for the effective refractive index, and $E_\text{X}=1.731\pm 0.003$~eV for the exciton energy in the metasurface region. The corresponding three-dimensional plot of the upper and lower polariton branches formed by the $Y$-polarized ($\pm1,0$) diffractive orders is shown in Fig.~\ref{fig2}e for the square lattice geometry defined in Fig.~\ref{fig2}f, and the respective two-dimensional projections in Figs.~\ref{fig2}g -- i. The fitting procedure yields a Rabi splitting $\Omega$ of $44\pm 5$~meV (see Methods for details) for the strongly coupled light-matter system, much larger than previously reported for polaritons in a TMD monolayer coupled to a dielectric metasurface \cite{chen2020metasurface}. Given the numbers for the competing coherent and incoherent processes, our polaritonic system clearly fulfills the common criteria for strong coupling, $g>|\gamma_{\text{X}}-\gamma_{\text{SLR}}|/2$ and $\Omega>\gamma_{\text{X}}+\gamma_{\text{SLR}}$ \cite{baranov2018novel,schneider2018two}.

With this quantitative understanding of the polaritonic van der Waals metasurface, we inspect its emission characteristics with momentum-resolved photoluminescence (PL) spectroscopy. The respective data in Fig.~\ref{fig3}a include contributions from both the strongly coupled monolayer excitons forming lattice polaritons and their uncoupled counterparts. To begin with the latter, we first note that the spectrum of the bare WSe$_2$ monolayer is known to be rich: in addition to the fundamental exciton X which strongly couples to SLRs to form the upper and lower polariton branches, we observe the PL of the biexciton XX \cite{you2015observation,barbone2018charge,li2018revealing,steinhoff2018biexciton} and the spin-forbidden dark exciton D \cite{zhang2015experimental,robert2017fine}, red-shifted from X by $19$ and $43$~meV, respectively. Additional contributions to zero-momentum dipolar PL include the negative trion doublet X$_1^-$ and X$_2^-$ with $6$~meV exchange splitting \cite{courtade2017charged}, visible in our sample due to residual doping. Finally, momentum-indirect excitons contribute with a series of PL sideband peaks labeled as M$_1$ through M$_3$. The assignment of M$_1$ at $33$~meV red-shift remains ambiguous, with direct PL emission from momentum-indirect $KK'$ \cite{he2020valley,liu2020multipath,yang2022relaxation} or a phonon sideband of $KQ$ excitons \cite{forste2020exciton} as possible origins. The peaks M$_2$ and M$_3$ at $50$ and $59$~meV red-shifts are consistently attributed to phonon sidebands of momentum-indirect $KK'$ excitons \cite{he2020valley,liu2020multipath}. 

To place emphasis on the emission characteristics of the polaritonic metasurface, we select an energy range of $4$~meV centered at $1.683$~eV. This energy range, slightly below the lower polariton maximum intensity, is void of any dipole-active transition of the bare monolayer. Note that even though M$_2$ is included in this spectral window, it has no means of coupling to the SLRs due to its momentum-indirect origin as phonon sideband with vanishingly small oscillator strength. Therefore, it contributes to the momentum-resolved PL characteristics its monolayer-specific background of out-of-plane dipolar emission, shown in Fig.~\ref{fig3}b and c. The structured emission profiles in Fig.~\ref{fig3}d and e, in contrast, are dominated by the characteristics of the metasurface lower polariton, exhibiting strong momentum narrowing and sensitivity to linear polarization. Under linear excitation and co-polarized detection, we observe a linearly polarized, directional far-field emission profile (Fig.~\ref{fig3}d), with a large fraction of the emission constrained to small angles of up to $12$° at maximum $k_y$. Under circularly polarized excitation and detection, both sets of linearly polarized branches with orthogonal orientation are observed simultaneously (Fig.~\ref{fig3}e). 

Clearly, the directional narrow-angle emission of the metasurface lower polariton is imprinted by the SLRs of the underlying nanoplasmonic square lattice (cf. Fig.~\ref{fig2}g). It also dictates the polarization characteristics of the coupled system, that is strongly modified as compared to the bare monolayer exciton transition. This effect is highlighted best by the direct comparison of the respective degrees of linear and circular PL polarization $P_\text{L}$ and $P_\text{C}$, determined from polarization-resolved PL as $P_\text{L/C}=(I_\text{co}-I_\text{cross})/(I_\text{co}+I_\text{cross})$, where $I_\text{co/cross}$ are the co/cross-polarized PL intensities. The monolayer WSe$_2$ exciton X is known to exhibit high degrees of valley coherence and polarization \cite{zeng2012valley,mak2012control,jones2013optical,wang2016control}, quantified by the respective degrees of linear and circular polarization of dipole-characteristic emission profile in Fig.~\ref{fig3}f and g. As the emission of the lower polariton branch is not dipolar but dictated by the underlying lattice, its degree of linear polarization in Fig.~\ref{fig3}h is highly structured (reaching values of up to $60\%$), as anticipated from the dispersions of the four lowest diffractive orders shown by the solid lines. The same dispersions are also evident in Fig.~\ref{fig3}i as lines of vanishing degree of circular polarization on a weak background of finite $P_\text{C}$ stemming from M$_2$ PL within the spectral window of integration.

In conclusion, our work establishes lattices of exciton-polaritons by strong coupling of the fundamental TMD monolayer exciton to the collective modes of nanoplasmonic arrays. The absorption of the strongly coupled system exhibits polariton characteristics imprinted by the lattice, substantiating the system as a polaritonic metasurface with narrow-angle, directional and linearly-polarized far-field emission profile. Replacing nanodiscs by chiral nanostructures would utilize the peculiar valleytronic properties of TMD monolayers for applications in opto-valleytronics \cite{gong2018nanoscale,hu2019coherent,guddala2019valley}. Clearly, our fabrication method, that ensures both the integrity of TMD monolayer exciton resonances and their coupling to the near-field of the plasmonic SLRs, is not limited to square lattices. It can be straightforwardly adapted to other lattice geometries to provide a complementary approach to the engineering of collective effects in purely photonic lattices \cite{Ozawa2019}, with a wide range of applications emulating exotic many-body phenomena such as Anderson localization \cite{Schwartz2007} or topological insulators \cite{Rechtsman2013}.

\section*{Methods}
\noindent \textbf{Sample fabrication}: The design of the gold nanodisk array with a lattice constant of $480$~nm, disk diameters of $70$~nm, and the height of $42$~nm matching the hBN thickness was optimized in simulations with the finite-difference time-domain method (software Lumerical). The thickness of the second hBN layer was optimized for spectral resonance of the SRL and the bright exciton X in monolayer WSe$_{2}$. High-quality WSe$_2$ monolayers were synthesized by in-house chemical vapor deposition \cite{li2022stacking}, and hBN flakes were exfoliated from bulk crystals (NIMS). Gold nanodisk arrays embedded in hBN were fabricated by employing electron-beam lithography (EBL) and inductively-coupled plasma (ICP) etching. A square array of nanodisks with a diameter of $70$~nm was defined via EBL using PMMA 950 K resist with a thickness of $200$~nm and a dose of $500$~$\mu$C/cm$^2$. Air-holes were etched into the hBN via ICP. The plasma was generated by $5$~sccm of Ar and $10$~sccm of SF$_{6}$ at a constant chamber pressure of $10$~mTorr. Setting the ICP power to $70$~W and the RF power to $6$~W yielded a controlled etch rate of $0.6$~nm/s. After etching, a gold film matching the thickness of the hBN layer was deposited via electron-beam evaporation. The hardened resist was lifted-off in acetone and isopropanol, and remaining residues were removed in an O$_{2}$ asher. 

For the assembly of the van der Waals metasurface, we used a dry stamping method with a polydimethylsiloxane (PDMS) droplet on a glass slide as stamp base, covered with a thin film of poly-bisphenol A-carbonate (PC, Sigma Aldrich) or polycaprolacton (PCL, Sigma Aldrich). We created the PC (PCL) film by dispersing a solution of polymer dissolved in chloroform (tetrahydrofurane, THF) at a mass percentage of 8\:\% (15\:\%). A $4$~nm bottom hBN flake was picked up with a PC stamp at $50$°C, released onto a glass substrate at $195$°C, and cleaned in chloroform, acetone, and isopropanol. For the pick-up of the hBN flake with the incorporated gold lattice, a PCL stamp was used. We started the process at 30$^{\circ}$C and heated up to 57$^{\circ}$C to cover the entire flake with the stamp. After cooling down to 30$^{\circ}$C, we lifted the flake with the embedded lattice off the substrate and released it onto a WSe$_{2}$ monolayer at 75$^{\circ}$C, followed by cleaning in THF, acetone, and isopropanol. Finally, a PC stamp was used to pick up the stack at $115$°C, followed by release onto the bottom hBN flake at $195$°C. The sample was cleaned in chloroform, acetone, and isopropanol and annealed in ultrahigh vacuum for $15$~h at $200$°C to remove trapped air and residues.

\vspace{8pt}
\noindent \textbf{Optical spectroscopy}: Cryogenic DR and PL spectroscopy were performed in back-scattering geometry in a closed-cycle cryostat (attocube systems, attoDRY800) with a base temperature of $4$~K. To position the sample with respect to a low temperature apochromatic objective (LT-APO/633-RAMAN/0.81), we used piezo-units (attocube systems, ANPx101, ANPz101, and ANSxy100). Angle-resolved DR and PL spectra were recorded with a lab-built Fourier imaging setup in 4f and telescope configuration employing four achromatic doublet lenses (Edmund Optics, VIS-NIR with focal lengths of $750$, $750$, $400$ and $150$~mm) and including a dove prism to rotate the Fourier image. The signal was dispersed by a monochromator (Teledyne Princeton Instruments, IsoPlane SCT320), with a 300 grooves/mm grating and detected by a Peltier-cooled CCD (Teledyne Princeton Instruments, PIXIS 1024). Narrow-energy momentum-space images were recorded by turning two tunable filters (Semrock 790 nm VersaChrome Edge tunable Shortpass and Longpass Filters) at zero-order spectrometer grating and an open slit. The conversion of camera pixel to wavevector in 1/$\mu$m was obtained via the size of the illuminated area on the CCD and using the relation $k_{\parallel}=k_{0}\text{sin}(\Theta)$ with the maximum angle $\Theta$ given by the numerical aperture of the objective, and $k_{0}=2\pi/\lambda$.  For both DR and PL measurements, we used a pulsed supercontinuum laser (NKT Photonics, SuperK with Varia). DR was defined as $DR = (R-R_0)/R_0$, where $R$ was the reflactance from the sample and $R_0$ was the reference reflectance from a region containing only hBN. PL was excited at $662\pm 5$~nm and $20~\mu$W.

\vspace{8pt}
\noindent \textbf{Plasmon-exciton-polariton dispersion}: The exciton-polariton dispersion was obtained from the coupled oscillator model \cite{savona1995quantum} with the Hamiltonian:
\begin{equation}
    H = 
    \begin{pmatrix}
        E_{\text{X}}-i\gamma_{\text{X}} & g \\
        g & E_{\text{SLR}}-i\gamma_{\text{SLR}}
    \end{pmatrix}.
\end{equation}
Diagonalization yields the well-known energy relation for the upper  and lower polariton branches given by Eq.~\eqref{ham} in the main text.  The Rabi splitting $\Omega$, defined as \cite{savona1995quantum}:
\begin{equation}\label{Rabi}
\Omega = 2 \sqrt{g^2-\frac{1}{4}(\gamma_\text{X}-\gamma_\text{SLR})^2},
\end{equation}
was determined experimentally as $44\pm5$~meV. The linewidth of the lower polariton branch is given by:
\begin{equation}\label{gamma_LP}
\gamma_\text{LP}=|X|^2 \gamma_\text{X} + |C|^2 \gamma_\text{SLR},
\end{equation}
where $X$ and $C$ are the Hopfield coefficients \cite{hopfield1958theory}:
\begin{align}\label{Hopf}
|X|^2 &= \frac{1}{2} \left[1+ \frac{E_\text{SLR}-E_\text{X}}{\sqrt{\Omega^2 + (E_\text{SLR}-E_\text{X})^2}} \right], \\
|C|^2 &= \frac{1}{2} \left[1- \frac{E_\text{SLR}-E_\text{X}}{\sqrt{\Omega^2 + (E_\text{SLR}-E_\text{X})^2}} \right],
\end{align}
and shown as dashed cyan lines in Figs.~\ref{fig2}c and \ref{fig3}a for the hyperbolic polariton dispersion.

\vspace{11pt}
\noindent \textbf{Author Information}

\vspace{6pt}
\noindent \textbf{Corresponding authors:}\\ F.\,T.-V. (f.tabataba@lmu.de) and A.\,H. (alexander.hoegele@lmu.de). 

\vspace{6pt}
\noindent \textbf{Author Contributions:}\\
F.\,T.-V. and L.\,K. contributed equally to this work.

\vspace{6pt}
\noindent \textbf{Notes:}\\
The authors declare no competing financial interest. 

\vspace{11pt}
\noindent \textbf{Acknowledgements}

\noindent This research was funded by the European Research Council (ERC) under the Grant Agreement No.~772195 and the Deutsche Forschungsgemeinschaft (DFG, German Research Foundation) within Germany's Excellence Strategy under Grant No.~EXC-2111-390814868. F.\,T.-V. acknowledges funding from the Munich Center for Quantum Science and Technology (MCQST) and the European Union's Framework Programme for Research and Innovation Horizon Europe under the Marie Sk{\l}odowska-Curie Actions grant agreement No.~101058981. K.\,W. and T.\,T. acknowledge support from the JSPS KAKENHI
(Grant Numbers 20H00354 and 23H02052) and World
Premier International Research Center Initiative (WPI),
MEXT, Japan. A.\,H. acknowledges funding by the Bavarian Hightech Agenda within the EQAP project.


\begin{thebibliography}{62}%
\makeatletter
\providecommand \@ifxundefined [1]{%
 \@ifx{#1\undefined}
}%
\providecommand \@ifnum [1]{%
 \ifnum #1\expandafter \@firstoftwo
 \else \expandafter \@secondoftwo
 \fi
}%
\providecommand \@ifx [1]{%
 \ifx #1\expandafter \@firstoftwo
 \else \expandafter \@secondoftwo
 \fi
}%
\providecommand \natexlab [1]{#1}%
\providecommand \enquote  [1]{``#1''}%
\providecommand \bibnamefont  [1]{#1}%
\providecommand \bibfnamefont [1]{#1}%
\providecommand \citenamefont [1]{#1}%
\providecommand \href@noop [0]{\@secondoftwo}%
\providecommand \href [0]{\begingroup \@sanitize@url \@href}%
\providecommand \@href[1]{\@@startlink{#1}\@@href}%
\providecommand \@@href[1]{\endgroup#1\@@endlink}%
\providecommand \@sanitize@url [0]{\catcode `\\12\catcode `\$12\catcode
  `\&12\catcode `\#12\catcode `\^12\catcode `\_12\catcode `\%12\relax}%
\providecommand \@@startlink[1]{}%
\providecommand \@@endlink[0]{}%
\providecommand \url  [0]{\begingroup\@sanitize@url \@url }%
\providecommand \@url [1]{\endgroup\@href {#1}{\urlprefix }}%
\providecommand \urlprefix  [0]{URL }%
\providecommand \Eprint [0]{\href }%
\providecommand \doibase [0]{https://doi.org/}%
\providecommand \selectlanguage [0]{\@gobble}%
\providecommand \bibinfo  [0]{\@secondoftwo}%
\providecommand \bibfield  [0]{\@secondoftwo}%
\providecommand \translation [1]{[#1]}%
\providecommand \BibitemOpen [0]{}%
\providecommand \bibitemStop [0]{}%
\providecommand \bibitemNoStop [0]{.\EOS\space}%
\providecommand \EOS [0]{\spacefactor3000\relax}%
\providecommand \BibitemShut  [1]{\csname bibitem#1\endcsname}%
\let\auto@bib@innerbib\@empty
\bibitem [{\citenamefont {Yu}\ and\ \citenamefont
  {Capasso}(2014)}]{yu2014flat}%
  \BibitemOpen
  \bibfield  {author} {\bibinfo {author} {\bibfnamefont {N.}~\bibnamefont
  {Yu}}\ and\ \bibinfo {author} {\bibfnamefont {F.}~\bibnamefont {Capasso}},\
  }\bibfield  {title} {\bibinfo {title} {Flat optics with designer
  metasurfaces},\ }\href@noop {} {\bibfield  {journal} {\bibinfo  {journal}
  {Nat. Mater.}\ }\textbf {\bibinfo {volume} {13}},\ \bibinfo {pages} {139}
  (\bibinfo {year} {2014})}\BibitemShut {NoStop}%
\bibitem [{\citenamefont {Meinzer}\ \emph {et~al.}(2014)\citenamefont
  {Meinzer}, \citenamefont {Barnes},\ and\ \citenamefont
  {Hooper}}]{meinzer2014plasmonic}%
  \BibitemOpen
  \bibfield  {author} {\bibinfo {author} {\bibfnamefont {N.}~\bibnamefont
  {Meinzer}}, \bibinfo {author} {\bibfnamefont {W.~L.}\ \bibnamefont
  {Barnes}},\ and\ \bibinfo {author} {\bibfnamefont {I.~R.}\ \bibnamefont
  {Hooper}},\ }\bibfield  {title} {\bibinfo {title} {Plasmonic meta-atoms and
  metasurfaces},\ }\href@noop {} {\bibfield  {journal} {\bibinfo  {journal}
  {Nat. Photonics}\ }\textbf {\bibinfo {volume} {8}},\ \bibinfo {pages} {889}
  (\bibinfo {year} {2014})}\BibitemShut {NoStop}%
\bibitem [{\citenamefont {Chen}\ \emph {et~al.}(2016)\citenamefont {Chen},
  \citenamefont {Taylor},\ and\ \citenamefont {Yu}}]{chen2016review}%
  \BibitemOpen
  \bibfield  {author} {\bibinfo {author} {\bibfnamefont {H.-T.}\ \bibnamefont
  {Chen}}, \bibinfo {author} {\bibfnamefont {A.~J.}\ \bibnamefont {Taylor}},\
  and\ \bibinfo {author} {\bibfnamefont {N.}~\bibnamefont {Yu}},\ }\bibfield
  {title} {\bibinfo {title} {A review of metasurfaces: physics and
  applications},\ }\href@noop {} {\bibfield  {journal} {\bibinfo  {journal}
  {Rep. Prog. Phys.}\ }\textbf {\bibinfo {volume} {79}},\ \bibinfo {pages}
  {076401} (\bibinfo {year} {2016})}\BibitemShut {NoStop}%
\bibitem [{\citenamefont {Yu}\ \emph {et~al.}(2011)\citenamefont {Yu},
  \citenamefont {Genevet}, \citenamefont {Kats}, \citenamefont {Aieta},
  \citenamefont {Tetienne}, \citenamefont {Capasso},\ and\ \citenamefont
  {Gaburro}}]{yu2011light}%
  \BibitemOpen
  \bibfield  {author} {\bibinfo {author} {\bibfnamefont {N.}~\bibnamefont
  {Yu}}, \bibinfo {author} {\bibfnamefont {P.}~\bibnamefont {Genevet}},
  \bibinfo {author} {\bibfnamefont {M.~A.}\ \bibnamefont {Kats}}, \bibinfo
  {author} {\bibfnamefont {F.}~\bibnamefont {Aieta}}, \bibinfo {author}
  {\bibfnamefont {J.-P.}\ \bibnamefont {Tetienne}}, \bibinfo {author}
  {\bibfnamefont {F.}~\bibnamefont {Capasso}},\ and\ \bibinfo {author}
  {\bibfnamefont {Z.}~\bibnamefont {Gaburro}},\ }\bibfield  {title} {\bibinfo
  {title} {Light propagation with phase discontinuities: generalized laws of
  reflection and refraction},\ }\href@noop {} {\bibfield  {journal} {\bibinfo
  {journal} {Science}\ }\textbf {\bibinfo {volume} {334}},\ \bibinfo {pages}
  {333} (\bibinfo {year} {2011})}\BibitemShut {NoStop}%
\bibitem [{\citenamefont {Huang}\ \emph {et~al.}(2013)\citenamefont {Huang},
  \citenamefont {Chen}, \citenamefont {M{\"u}hlenbernd}, \citenamefont {Zhang},
  \citenamefont {Chen}, \citenamefont {Bai}, \citenamefont {Tan}, \citenamefont
  {Jin}, \citenamefont {Cheah}, \citenamefont {Qiu}, \citenamefont {Li},
  \citenamefont {Zentgraf},\ and\ \citenamefont {Zhang}}]{huang2013three}%
  \BibitemOpen
  \bibfield  {author} {\bibinfo {author} {\bibfnamefont {L.}~\bibnamefont
  {Huang}}, \bibinfo {author} {\bibfnamefont {X.}~\bibnamefont {Chen}},
  \bibinfo {author} {\bibfnamefont {H.}~\bibnamefont {M{\"u}hlenbernd}},
  \bibinfo {author} {\bibfnamefont {H.}~\bibnamefont {Zhang}}, \bibinfo
  {author} {\bibfnamefont {S.}~\bibnamefont {Chen}}, \bibinfo {author}
  {\bibfnamefont {B.}~\bibnamefont {Bai}}, \bibinfo {author} {\bibfnamefont
  {Q.}~\bibnamefont {Tan}}, \bibinfo {author} {\bibfnamefont {G.}~\bibnamefont
  {Jin}}, \bibinfo {author} {\bibfnamefont {K.-W.}\ \bibnamefont {Cheah}},
  \bibinfo {author} {\bibfnamefont {C.-W.}\ \bibnamefont {Qiu}}, \bibinfo
  {author} {\bibfnamefont {J.}~\bibnamefont {Li}}, \bibinfo {author}
  {\bibfnamefont {T.}~\bibnamefont {Zentgraf}},\ and\ \bibinfo {author}
  {\bibfnamefont {S.}~\bibnamefont {Zhang}},\ }\bibfield  {title} {\bibinfo
  {title} {Three-dimensional optical holography using a plasmonic
  metasurface},\ }\href@noop {} {\bibfield  {journal} {\bibinfo  {journal}
  {Nat. Commun.}\ }\textbf {\bibinfo {volume} {4}},\ \bibinfo {pages} {2808}
  (\bibinfo {year} {2013})}\BibitemShut {NoStop}%
\bibitem [{\citenamefont {Aieta}\ \emph {et~al.}(2012)\citenamefont {Aieta},
  \citenamefont {Genevet}, \citenamefont {Kats}, \citenamefont {Yu},
  \citenamefont {Blanchard}, \citenamefont {Gaburro},\ and\ \citenamefont
  {Capasso}}]{aieta2012aberration}%
  \BibitemOpen
  \bibfield  {author} {\bibinfo {author} {\bibfnamefont {F.}~\bibnamefont
  {Aieta}}, \bibinfo {author} {\bibfnamefont {P.}~\bibnamefont {Genevet}},
  \bibinfo {author} {\bibfnamefont {M.~A.}\ \bibnamefont {Kats}}, \bibinfo
  {author} {\bibfnamefont {N.}~\bibnamefont {Yu}}, \bibinfo {author}
  {\bibfnamefont {R.}~\bibnamefont {Blanchard}}, \bibinfo {author}
  {\bibfnamefont {Z.}~\bibnamefont {Gaburro}},\ and\ \bibinfo {author}
  {\bibfnamefont {F.}~\bibnamefont {Capasso}},\ }\bibfield  {title} {\bibinfo
  {title} {Aberration-free ultrathin flat lenses and axicons at telecom
  wavelengths based on plasmonic metasurfaces},\ }\href@noop {} {\bibfield
  {journal} {\bibinfo  {journal} {Nano Lett.}\ }\textbf {\bibinfo {volume}
  {12}},\ \bibinfo {pages} {4932} (\bibinfo {year} {2012})}\BibitemShut
  {NoStop}%
\bibitem [{\citenamefont {Deng}\ \emph {et~al.}(2020)\citenamefont {Deng},
  \citenamefont {Li}, \citenamefont {Park}, \citenamefont {Guan}, \citenamefont
  {Choo}, \citenamefont {Hu}, \citenamefont {Smeets},\ and\ \citenamefont
  {Odom}}]{deng2020ultranarrow}%
  \BibitemOpen
  \bibfield  {author} {\bibinfo {author} {\bibfnamefont {S.}~\bibnamefont
  {Deng}}, \bibinfo {author} {\bibfnamefont {R.}~\bibnamefont {Li}}, \bibinfo
  {author} {\bibfnamefont {J.-E.}\ \bibnamefont {Park}}, \bibinfo {author}
  {\bibfnamefont {J.}~\bibnamefont {Guan}}, \bibinfo {author} {\bibfnamefont
  {P.}~\bibnamefont {Choo}}, \bibinfo {author} {\bibfnamefont {J.}~\bibnamefont
  {Hu}}, \bibinfo {author} {\bibfnamefont {P.~J.}\ \bibnamefont {Smeets}},\
  and\ \bibinfo {author} {\bibfnamefont {T.~W.}\ \bibnamefont {Odom}},\
  }\bibfield  {title} {\bibinfo {title} {Ultranarrow plasmon resonances from
  annealed nanoparticle lattices},\ }\href@noop {} {\bibfield  {journal}
  {\bibinfo  {journal} {Proc. Natl. Acad. Sci. U.S.A.}\ }\textbf {\bibinfo
  {volume} {117}},\ \bibinfo {pages} {23380} (\bibinfo {year}
  {2020})}\BibitemShut {NoStop}%
\bibitem [{\citenamefont {Cherqui}\ \emph {et~al.}(2019)\citenamefont
  {Cherqui}, \citenamefont {Bourgeois}, \citenamefont {Wang},\ and\
  \citenamefont {Schatz}}]{cherqui2019plasmonic}%
  \BibitemOpen
  \bibfield  {author} {\bibinfo {author} {\bibfnamefont {C.}~\bibnamefont
  {Cherqui}}, \bibinfo {author} {\bibfnamefont {M.~R.}\ \bibnamefont
  {Bourgeois}}, \bibinfo {author} {\bibfnamefont {D.}~\bibnamefont {Wang}},\
  and\ \bibinfo {author} {\bibfnamefont {G.~C.}\ \bibnamefont {Schatz}},\
  }\bibfield  {title} {\bibinfo {title} {Plasmonic surface lattice resonances:
  Theory and computation},\ }\href@noop {} {\bibfield  {journal} {\bibinfo
  {journal} {Acc. Chem. Res.}\ }\textbf {\bibinfo {volume} {52}},\ \bibinfo
  {pages} {2548} (\bibinfo {year} {2019})}\BibitemShut {NoStop}%
\bibitem [{\citenamefont {Guo}\ \emph {et~al.}(2017)\citenamefont {Guo},
  \citenamefont {Hakala},\ and\ \citenamefont
  {T{\"o}rm{\"a}}}]{guo2017geometry}%
  \BibitemOpen
  \bibfield  {author} {\bibinfo {author} {\bibfnamefont {R.}~\bibnamefont
  {Guo}}, \bibinfo {author} {\bibfnamefont {T.~K.}\ \bibnamefont {Hakala}},\
  and\ \bibinfo {author} {\bibfnamefont {P.}~\bibnamefont {T{\"o}rm{\"a}}},\
  }\bibfield  {title} {\bibinfo {title} {Geometry dependence of surface lattice
  resonances in plasmonic nanoparticle arrays},\ }\href@noop {} {\bibfield
  {journal} {\bibinfo  {journal} {Phys. Rev. B}\ }\textbf {\bibinfo {volume}
  {95}},\ \bibinfo {pages} {155423} (\bibinfo {year} {2017})}\BibitemShut
  {NoStop}%
\bibitem [{\citenamefont {Meng}\ \emph {et~al.}(2023)\citenamefont {Meng},
  \citenamefont {Zhong}, \citenamefont {Xu}, \citenamefont {He}, \citenamefont
  {Kim}, \citenamefont {Han}, \citenamefont {Kim}, \citenamefont {Park},
  \citenamefont {Shen}, \citenamefont {Gong}, \citenamefont {Xiao},\ and\
  \citenamefont {Bae}}]{meng2023functionalizing}%
  \BibitemOpen
  \bibfield  {author} {\bibinfo {author} {\bibfnamefont {Y.}~\bibnamefont
  {Meng}}, \bibinfo {author} {\bibfnamefont {H.}~\bibnamefont {Zhong}},
  \bibinfo {author} {\bibfnamefont {Z.}~\bibnamefont {Xu}}, \bibinfo {author}
  {\bibfnamefont {T.}~\bibnamefont {He}}, \bibinfo {author} {\bibfnamefont
  {J.~S.}\ \bibnamefont {Kim}}, \bibinfo {author} {\bibfnamefont
  {S.}~\bibnamefont {Han}}, \bibinfo {author} {\bibfnamefont {S.}~\bibnamefont
  {Kim}}, \bibinfo {author} {\bibfnamefont {S.}~\bibnamefont {Park}}, \bibinfo
  {author} {\bibfnamefont {Y.}~\bibnamefont {Shen}}, \bibinfo {author}
  {\bibfnamefont {M.}~\bibnamefont {Gong}}, \bibinfo {author} {\bibfnamefont
  {Q.}~\bibnamefont {Xiao}},\ and\ \bibinfo {author} {\bibfnamefont {S.-H.}\
  \bibnamefont {Bae}},\ }\bibfield  {title} {\bibinfo {title} {Functionalizing
  nanophotonic structures with {2D} van der {W}aals materials},\ }\href@noop {}
  {\bibfield  {journal} {\bibinfo  {journal} {Nanoscale Horiz.}\ }\textbf
  {\bibinfo {volume} {8}},\ \bibinfo {pages} {1345} (\bibinfo {year}
  {2023})}\BibitemShut {NoStop}%
\bibitem [{\citenamefont {Pizzocchero}\ \emph {et~al.}(2016)\citenamefont
  {Pizzocchero}, \citenamefont {Gammelgaard}, \citenamefont {Jessen},
  \citenamefont {Caridad}, \citenamefont {Wang}, \citenamefont {Hone},
  \citenamefont {B{\o}ggild},\ and\ \citenamefont
  {Booth}}]{pizzocchero2016hot}%
  \BibitemOpen
  \bibfield  {author} {\bibinfo {author} {\bibfnamefont {F.}~\bibnamefont
  {Pizzocchero}}, \bibinfo {author} {\bibfnamefont {L.}~\bibnamefont
  {Gammelgaard}}, \bibinfo {author} {\bibfnamefont {B.~S.}\ \bibnamefont
  {Jessen}}, \bibinfo {author} {\bibfnamefont {J.~M.}\ \bibnamefont {Caridad}},
  \bibinfo {author} {\bibfnamefont {L.}~\bibnamefont {Wang}}, \bibinfo {author}
  {\bibfnamefont {J.}~\bibnamefont {Hone}}, \bibinfo {author} {\bibfnamefont
  {P.}~\bibnamefont {B{\o}ggild}},\ and\ \bibinfo {author} {\bibfnamefont
  {T.~J.}\ \bibnamefont {Booth}},\ }\bibfield  {title} {\bibinfo {title} {The
  hot pick-up technique for batch assembly of van der {W}aals
  heterostructures},\ }\href@noop {} {\bibfield  {journal} {\bibinfo  {journal}
  {Nat. Commun.}\ }\textbf {\bibinfo {volume} {7}},\ \bibinfo {pages} {11894}
  (\bibinfo {year} {2016})}\BibitemShut {NoStop}%
\bibitem [{\citenamefont {Splendiani}\ \emph {et~al.}(2010)\citenamefont
  {Splendiani}, \citenamefont {Sun}, \citenamefont {Zhang}, \citenamefont {Li},
  \citenamefont {Kim}, \citenamefont {Chim}, \citenamefont {Galli},\ and\
  \citenamefont {Wang}}]{splendiani2010emerging}%
  \BibitemOpen
  \bibfield  {author} {\bibinfo {author} {\bibfnamefont {A.}~\bibnamefont
  {Splendiani}}, \bibinfo {author} {\bibfnamefont {L.}~\bibnamefont {Sun}},
  \bibinfo {author} {\bibfnamefont {Y.}~\bibnamefont {Zhang}}, \bibinfo
  {author} {\bibfnamefont {T.}~\bibnamefont {Li}}, \bibinfo {author}
  {\bibfnamefont {J.}~\bibnamefont {Kim}}, \bibinfo {author} {\bibfnamefont
  {C.-Y.}\ \bibnamefont {Chim}}, \bibinfo {author} {\bibfnamefont
  {G.}~\bibnamefont {Galli}},\ and\ \bibinfo {author} {\bibfnamefont
  {F.}~\bibnamefont {Wang}},\ }\bibfield  {title} {\bibinfo {title} {Emerging
  photoluminescence in monolayer {MoS}$_2$},\ }\href@noop {} {\bibfield
  {journal} {\bibinfo  {journal} {Nano Lett.}\ }\textbf {\bibinfo {volume}
  {10}},\ \bibinfo {pages} {1271} (\bibinfo {year} {2010})}\BibitemShut
  {NoStop}%
\bibitem [{\citenamefont {Mak}\ \emph {et~al.}(2010)\citenamefont {Mak},
  \citenamefont {Lee}, \citenamefont {Hone}, \citenamefont {Shan},\ and\
  \citenamefont {Heinz}}]{mak2010atomically}%
  \BibitemOpen
  \bibfield  {author} {\bibinfo {author} {\bibfnamefont {K.~F.}\ \bibnamefont
  {Mak}}, \bibinfo {author} {\bibfnamefont {C.}~\bibnamefont {Lee}}, \bibinfo
  {author} {\bibfnamefont {J.}~\bibnamefont {Hone}}, \bibinfo {author}
  {\bibfnamefont {J.}~\bibnamefont {Shan}},\ and\ \bibinfo {author}
  {\bibfnamefont {T.~F.}\ \bibnamefont {Heinz}},\ }\bibfield  {title} {\bibinfo
  {title} {Atomically thin {MoS}$_2$: a new direct-gap semiconductor},\
  }\href@noop {} {\bibfield  {journal} {\bibinfo  {journal} {Phys. Rev. Lett.}\
  }\textbf {\bibinfo {volume} {105}},\ \bibinfo {pages} {136805} (\bibinfo
  {year} {2010})}\BibitemShut {NoStop}%
\bibitem [{\citenamefont {He}\ \emph {et~al.}(2014)\citenamefont {He},
  \citenamefont {Kumar}, \citenamefont {Zhao}, \citenamefont {Wang},
  \citenamefont {Mak}, \citenamefont {Zhao},\ and\ \citenamefont
  {Shan}}]{he2014tightly}%
  \BibitemOpen
  \bibfield  {author} {\bibinfo {author} {\bibfnamefont {K.}~\bibnamefont
  {He}}, \bibinfo {author} {\bibfnamefont {N.}~\bibnamefont {Kumar}}, \bibinfo
  {author} {\bibfnamefont {L.}~\bibnamefont {Zhao}}, \bibinfo {author}
  {\bibfnamefont {Z.}~\bibnamefont {Wang}}, \bibinfo {author} {\bibfnamefont
  {K.~F.}\ \bibnamefont {Mak}}, \bibinfo {author} {\bibfnamefont
  {H.}~\bibnamefont {Zhao}},\ and\ \bibinfo {author} {\bibfnamefont
  {J.}~\bibnamefont {Shan}},\ }\bibfield  {title} {\bibinfo {title} {Tightly
  bound excitons in monolayer {WSe}$_2$},\ }\href@noop {} {\bibfield  {journal}
  {\bibinfo  {journal} {Phys. Rev. Lett.}\ }\textbf {\bibinfo {volume} {113}},\
  \bibinfo {pages} {026803} (\bibinfo {year} {2014})}\BibitemShut {NoStop}%
\bibitem [{\citenamefont {Chernikov}\ \emph {et~al.}(2014)\citenamefont
  {Chernikov}, \citenamefont {Berkelbach}, \citenamefont {Hill}, \citenamefont
  {Rigosi}, \citenamefont {Li}, \citenamefont {Aslan}, \citenamefont
  {Reichman}, \citenamefont {Hybertsen},\ and\ \citenamefont
  {Heinz}}]{chernikov2014exciton}%
  \BibitemOpen
  \bibfield  {author} {\bibinfo {author} {\bibfnamefont {A.}~\bibnamefont
  {Chernikov}}, \bibinfo {author} {\bibfnamefont {T.~C.}\ \bibnamefont
  {Berkelbach}}, \bibinfo {author} {\bibfnamefont {H.~M.}\ \bibnamefont
  {Hill}}, \bibinfo {author} {\bibfnamefont {A.}~\bibnamefont {Rigosi}},
  \bibinfo {author} {\bibfnamefont {Y.}~\bibnamefont {Li}}, \bibinfo {author}
  {\bibfnamefont {B.}~\bibnamefont {Aslan}}, \bibinfo {author} {\bibfnamefont
  {D.~R.}\ \bibnamefont {Reichman}}, \bibinfo {author} {\bibfnamefont {M.~S.}\
  \bibnamefont {Hybertsen}},\ and\ \bibinfo {author} {\bibfnamefont {T.~F.}\
  \bibnamefont {Heinz}},\ }\bibfield  {title} {\bibinfo {title} {Exciton
  binding energy and nonhydrogenic {R}ydberg series in monolayer {WS}$_2$},\
  }\href@noop {} {\bibfield  {journal} {\bibinfo  {journal} {Phys. Rev. Lett.}\
  }\textbf {\bibinfo {volume} {113}},\ \bibinfo {pages} {076802} (\bibinfo
  {year} {2014})}\BibitemShut {NoStop}%
\bibitem [{\citenamefont {Zhang}\ \emph {et~al.}(2014)\citenamefont {Zhang},
  \citenamefont {Wang}, \citenamefont {Chan}, \citenamefont {Manolatou},\ and\
  \citenamefont {Rana}}]{zhang2014absorption}%
  \BibitemOpen
  \bibfield  {author} {\bibinfo {author} {\bibfnamefont {C.}~\bibnamefont
  {Zhang}}, \bibinfo {author} {\bibfnamefont {H.}~\bibnamefont {Wang}},
  \bibinfo {author} {\bibfnamefont {W.}~\bibnamefont {Chan}}, \bibinfo {author}
  {\bibfnamefont {C.}~\bibnamefont {Manolatou}},\ and\ \bibinfo {author}
  {\bibfnamefont {F.}~\bibnamefont {Rana}},\ }\bibfield  {title} {\bibinfo
  {title} {Absorption of light by excitons and trions in monolayers of metal
  dichalcogenide {MoS}$_2$: Experiments and theory},\ }\href@noop {} {\bibfield
   {journal} {\bibinfo  {journal} {Phys. Rev. B}\ }\textbf {\bibinfo {volume}
  {89}},\ \bibinfo {pages} {205436} (\bibinfo {year} {2014})}\BibitemShut
  {NoStop}%
\bibitem [{\citenamefont {Li}\ \emph {et~al.}(2014)\citenamefont {Li},
  \citenamefont {Chernikov}, \citenamefont {Zhang}, \citenamefont {Rigosi},
  \citenamefont {Hill}, \citenamefont {Van Der~Zande}, \citenamefont {Chenet},
  \citenamefont {Shih}, \citenamefont {Hone},\ and\ \citenamefont
  {Heinz}}]{li2014measurement}%
  \BibitemOpen
  \bibfield  {author} {\bibinfo {author} {\bibfnamefont {Y.}~\bibnamefont
  {Li}}, \bibinfo {author} {\bibfnamefont {A.}~\bibnamefont {Chernikov}},
  \bibinfo {author} {\bibfnamefont {X.}~\bibnamefont {Zhang}}, \bibinfo
  {author} {\bibfnamefont {A.}~\bibnamefont {Rigosi}}, \bibinfo {author}
  {\bibfnamefont {H.~M.}\ \bibnamefont {Hill}}, \bibinfo {author}
  {\bibfnamefont {A.~M.}\ \bibnamefont {Van Der~Zande}}, \bibinfo {author}
  {\bibfnamefont {D.~A.}\ \bibnamefont {Chenet}}, \bibinfo {author}
  {\bibfnamefont {E.-M.}\ \bibnamefont {Shih}}, \bibinfo {author}
  {\bibfnamefont {J.}~\bibnamefont {Hone}},\ and\ \bibinfo {author}
  {\bibfnamefont {T.~F.}\ \bibnamefont {Heinz}},\ }\bibfield  {title} {\bibinfo
  {title} {Measurement of the optical dielectric function of monolayer
  transition-metal dichalcogenides: {MoS}$_2$, {MoSe}$_2$, {WS}$_2$, and
  {WSe}$_2$},\ }\href@noop {} {\bibfield  {journal} {\bibinfo  {journal} {Phys.
  Rev. B}\ }\textbf {\bibinfo {volume} {90}},\ \bibinfo {pages} {205422}
  (\bibinfo {year} {2014})}\BibitemShut {NoStop}%
\bibitem [{\citenamefont {Xiao}\ \emph {et~al.}(2012)\citenamefont {Xiao},
  \citenamefont {Liu}, \citenamefont {Feng}, \citenamefont {Xu},\ and\
  \citenamefont {Yao}}]{xiao2012coupled}%
  \BibitemOpen
  \bibfield  {author} {\bibinfo {author} {\bibfnamefont {D.}~\bibnamefont
  {Xiao}}, \bibinfo {author} {\bibfnamefont {G.-B.}\ \bibnamefont {Liu}},
  \bibinfo {author} {\bibfnamefont {W.}~\bibnamefont {Feng}}, \bibinfo {author}
  {\bibfnamefont {X.}~\bibnamefont {Xu}},\ and\ \bibinfo {author}
  {\bibfnamefont {W.}~\bibnamefont {Yao}},\ }\bibfield  {title} {\bibinfo
  {title} {Coupled spin and valley physics in monolayers of {MoS}$_2$ and other
  group-{VI} dichalcogenides},\ }\href@noop {} {\bibfield  {journal} {\bibinfo
  {journal} {Phys. Rev. Lett.}\ }\textbf {\bibinfo {volume} {108}},\ \bibinfo
  {pages} {196802} (\bibinfo {year} {2012})}\BibitemShut {NoStop}%
\bibitem [{\citenamefont {Mak}\ \emph {et~al.}(2012)\citenamefont {Mak},
  \citenamefont {He}, \citenamefont {Shan},\ and\ \citenamefont
  {Heinz}}]{mak2012control}%
  \BibitemOpen
  \bibfield  {author} {\bibinfo {author} {\bibfnamefont {K.~F.}\ \bibnamefont
  {Mak}}, \bibinfo {author} {\bibfnamefont {K.}~\bibnamefont {He}}, \bibinfo
  {author} {\bibfnamefont {J.}~\bibnamefont {Shan}},\ and\ \bibinfo {author}
  {\bibfnamefont {T.~F.}\ \bibnamefont {Heinz}},\ }\bibfield  {title} {\bibinfo
  {title} {Control of valley polarization in monolayer {MoS}$_2$ by optical
  helicity},\ }\href@noop {} {\bibfield  {journal} {\bibinfo  {journal} {Nat.
  Nanotechnol.}\ }\textbf {\bibinfo {volume} {7}},\ \bibinfo {pages} {494}
  (\bibinfo {year} {2012})}\BibitemShut {NoStop}%
\bibitem [{\citenamefont {Wang}\ \emph {et~al.}(2018)\citenamefont {Wang},
  \citenamefont {Chernikov}, \citenamefont {Glazov}, \citenamefont {Heinz},
  \citenamefont {Marie}, \citenamefont {Amand},\ and\ \citenamefont
  {Urbaszek}}]{wang2018colloquium}%
  \BibitemOpen
  \bibfield  {author} {\bibinfo {author} {\bibfnamefont {G.}~\bibnamefont
  {Wang}}, \bibinfo {author} {\bibfnamefont {A.}~\bibnamefont {Chernikov}},
  \bibinfo {author} {\bibfnamefont {M.~M.}\ \bibnamefont {Glazov}}, \bibinfo
  {author} {\bibfnamefont {T.~F.}\ \bibnamefont {Heinz}}, \bibinfo {author}
  {\bibfnamefont {X.}~\bibnamefont {Marie}}, \bibinfo {author} {\bibfnamefont
  {T.}~\bibnamefont {Amand}},\ and\ \bibinfo {author} {\bibfnamefont
  {B.}~\bibnamefont {Urbaszek}},\ }\bibfield  {title} {\bibinfo {title}
  {Colloquium: Excitons in atomically thin transition metal dichalcogenides},\
  }\href@noop {} {\bibfield  {journal} {\bibinfo  {journal} {Rev. Mod. Phys.}\
  }\textbf {\bibinfo {volume} {90}},\ \bibinfo {pages} {021001} (\bibinfo
  {year} {2018})}\BibitemShut {NoStop}%
\bibitem [{\citenamefont {Van~de Groep}\ \emph {et~al.}(2020)\citenamefont
  {Van~de Groep}, \citenamefont {Song}, \citenamefont {Celano}, \citenamefont
  {Li}, \citenamefont {Kik},\ and\ \citenamefont
  {Brongersma}}]{van2020exciton}%
  \BibitemOpen
  \bibfield  {author} {\bibinfo {author} {\bibfnamefont {J.}~\bibnamefont
  {Van~de Groep}}, \bibinfo {author} {\bibfnamefont {J.-H.}\ \bibnamefont
  {Song}}, \bibinfo {author} {\bibfnamefont {U.}~\bibnamefont {Celano}},
  \bibinfo {author} {\bibfnamefont {Q.}~\bibnamefont {Li}}, \bibinfo {author}
  {\bibfnamefont {P.~G.}\ \bibnamefont {Kik}},\ and\ \bibinfo {author}
  {\bibfnamefont {M.~L.}\ \bibnamefont {Brongersma}},\ }\bibfield  {title}
  {\bibinfo {title} {Exciton resonance tuning of an atomically thin lens},\
  }\href@noop {} {\bibfield  {journal} {\bibinfo  {journal} {Nat. Photonics}\
  }\textbf {\bibinfo {volume} {14}},\ \bibinfo {pages} {426} (\bibinfo {year}
  {2020})}\BibitemShut {NoStop}%
\bibitem [{\citenamefont {Sun}\ \emph {et~al.}(2019)\citenamefont {Sun},
  \citenamefont {Wang}, \citenamefont {Krasnok}, \citenamefont {Choi},
  \citenamefont {Shi}, \citenamefont {Gomez-Diaz}, \citenamefont {Zepeda},
  \citenamefont {Gwo}, \citenamefont {Shih}, \citenamefont {Al{\`u}},\ and\
  \citenamefont {Li}}]{sun2019separation}%
  \BibitemOpen
  \bibfield  {author} {\bibinfo {author} {\bibfnamefont {L.}~\bibnamefont
  {Sun}}, \bibinfo {author} {\bibfnamefont {C.-Y.}\ \bibnamefont {Wang}},
  \bibinfo {author} {\bibfnamefont {A.}~\bibnamefont {Krasnok}}, \bibinfo
  {author} {\bibfnamefont {J.}~\bibnamefont {Choi}}, \bibinfo {author}
  {\bibfnamefont {J.}~\bibnamefont {Shi}}, \bibinfo {author} {\bibfnamefont
  {J.~S.}\ \bibnamefont {Gomez-Diaz}}, \bibinfo {author} {\bibfnamefont
  {A.}~\bibnamefont {Zepeda}}, \bibinfo {author} {\bibfnamefont
  {S.}~\bibnamefont {Gwo}}, \bibinfo {author} {\bibfnamefont {C.-K.}\
  \bibnamefont {Shih}}, \bibinfo {author} {\bibfnamefont {A.}~\bibnamefont
  {Al{\`u}}},\ and\ \bibinfo {author} {\bibfnamefont {X.}~\bibnamefont {Li}},\
  }\bibfield  {title} {\bibinfo {title} {Separation of valley excitons in a
  {MoS}$_2$ monolayer using a subwavelength asymmetric groove array},\
  }\href@noop {} {\bibfield  {journal} {\bibinfo  {journal} {Nat. Photonics}\
  }\textbf {\bibinfo {volume} {13}},\ \bibinfo {pages} {180} (\bibinfo {year}
  {2019})}\BibitemShut {NoStop}%
\bibitem [{\citenamefont {Dufferwiel}\ \emph {et~al.}(2015)\citenamefont
  {Dufferwiel}, \citenamefont {Schwarz}, \citenamefont {Withers}, \citenamefont
  {Trichet}, \citenamefont {Li}, \citenamefont {Sich}, \citenamefont {Del
  Pozo-Zamudio}, \citenamefont {Clark}, \citenamefont {Nalitov}, \citenamefont
  {Solnyshkov}, \citenamefont {Malpuech}, \citenamefont {Novoselov},
  \citenamefont {Smith}, \citenamefont {Skolnick}, \citenamefont
  {Krizhanovskii},\ and\ \citenamefont {Tartakovskii}}]{dufferwiel2015exciton}%
  \BibitemOpen
  \bibfield  {author} {\bibinfo {author} {\bibfnamefont {S.}~\bibnamefont
  {Dufferwiel}}, \bibinfo {author} {\bibfnamefont {S.}~\bibnamefont {Schwarz}},
  \bibinfo {author} {\bibfnamefont {F.}~\bibnamefont {Withers}}, \bibinfo
  {author} {\bibfnamefont {A.}~\bibnamefont {Trichet}}, \bibinfo {author}
  {\bibfnamefont {F.}~\bibnamefont {Li}}, \bibinfo {author} {\bibfnamefont
  {M.}~\bibnamefont {Sich}}, \bibinfo {author} {\bibfnamefont {O.}~\bibnamefont
  {Del Pozo-Zamudio}}, \bibinfo {author} {\bibfnamefont {C.}~\bibnamefont
  {Clark}}, \bibinfo {author} {\bibfnamefont {A.}~\bibnamefont {Nalitov}},
  \bibinfo {author} {\bibfnamefont {D.}~\bibnamefont {Solnyshkov}}, \bibinfo
  {author} {\bibfnamefont {G.}~\bibnamefont {Malpuech}}, \bibinfo {author}
  {\bibfnamefont {K.~S.}\ \bibnamefont {Novoselov}}, \bibinfo {author}
  {\bibfnamefont {J.~M.}\ \bibnamefont {Smith}}, \bibinfo {author}
  {\bibfnamefont {M.~S.}\ \bibnamefont {Skolnick}}, \bibinfo {author}
  {\bibfnamefont {D.~N.}\ \bibnamefont {Krizhanovskii}},\ and\ \bibinfo
  {author} {\bibfnamefont {A.~I.}\ \bibnamefont {Tartakovskii}},\ }\bibfield
  {title} {\bibinfo {title} {Exciton--polaritons in van der {W}aals
  heterostructures embedded in tunable microcavities},\ }\href@noop {}
  {\bibfield  {journal} {\bibinfo  {journal} {Nat. Commun.}\ }\textbf {\bibinfo
  {volume} {6}},\ \bibinfo {pages} {8579} (\bibinfo {year} {2015})}\BibitemShut
  {NoStop}%
\bibitem [{\citenamefont {Liu}\ \emph {et~al.}(2015)\citenamefont {Liu},
  \citenamefont {Galfsky}, \citenamefont {Sun}, \citenamefont {Xia},
  \citenamefont {Lin}, \citenamefont {Lee}, \citenamefont {K{\'e}na-Cohen},\
  and\ \citenamefont {Menon}}]{liu2015strong}%
  \BibitemOpen
  \bibfield  {author} {\bibinfo {author} {\bibfnamefont {X.}~\bibnamefont
  {Liu}}, \bibinfo {author} {\bibfnamefont {T.}~\bibnamefont {Galfsky}},
  \bibinfo {author} {\bibfnamefont {Z.}~\bibnamefont {Sun}}, \bibinfo {author}
  {\bibfnamefont {F.}~\bibnamefont {Xia}}, \bibinfo {author} {\bibfnamefont
  {E.-c.}\ \bibnamefont {Lin}}, \bibinfo {author} {\bibfnamefont {Y.-H.}\
  \bibnamefont {Lee}}, \bibinfo {author} {\bibfnamefont {S.}~\bibnamefont
  {K{\'e}na-Cohen}},\ and\ \bibinfo {author} {\bibfnamefont {V.~M.}\
  \bibnamefont {Menon}},\ }\bibfield  {title} {\bibinfo {title} {Strong
  light--matter coupling in two-dimensional atomic crystals},\ }\href@noop {}
  {\bibfield  {journal} {\bibinfo  {journal} {Nat. Photonics}\ }\textbf
  {\bibinfo {volume} {9}},\ \bibinfo {pages} {30} (\bibinfo {year}
  {2015})}\BibitemShut {NoStop}%
\bibitem [{\citenamefont {Chen}\ \emph {et~al.}(2020)\citenamefont {Chen},
  \citenamefont {Miao}, \citenamefont {Wang}, \citenamefont {Zhong},
  \citenamefont {Saxena}, \citenamefont {Chow}, \citenamefont {Whitehead},
  \citenamefont {Gerace}, \citenamefont {Xu}, \citenamefont {Shi},\ and\
  \citenamefont {Majumdar}}]{chen2020metasurface}%
  \BibitemOpen
  \bibfield  {author} {\bibinfo {author} {\bibfnamefont {Y.}~\bibnamefont
  {Chen}}, \bibinfo {author} {\bibfnamefont {S.}~\bibnamefont {Miao}}, \bibinfo
  {author} {\bibfnamefont {T.}~\bibnamefont {Wang}}, \bibinfo {author}
  {\bibfnamefont {D.}~\bibnamefont {Zhong}}, \bibinfo {author} {\bibfnamefont
  {A.}~\bibnamefont {Saxena}}, \bibinfo {author} {\bibfnamefont
  {C.}~\bibnamefont {Chow}}, \bibinfo {author} {\bibfnamefont {J.}~\bibnamefont
  {Whitehead}}, \bibinfo {author} {\bibfnamefont {D.}~\bibnamefont {Gerace}},
  \bibinfo {author} {\bibfnamefont {X.}~\bibnamefont {Xu}}, \bibinfo {author}
  {\bibfnamefont {S.-F.}\ \bibnamefont {Shi}},\ and\ \bibinfo {author}
  {\bibfnamefont {A.}~\bibnamefont {Majumdar}},\ }\bibfield  {title} {\bibinfo
  {title} {Metasurface integrated monolayer exciton polariton},\ }\href@noop {}
  {\bibfield  {journal} {\bibinfo  {journal} {Nano Lett.}\ }\textbf {\bibinfo
  {volume} {20}},\ \bibinfo {pages} {5292} (\bibinfo {year}
  {2020})}\BibitemShut {NoStop}%
\bibitem [{\citenamefont {Liu}\ \emph {et~al.}(2016)\citenamefont {Liu},
  \citenamefont {Lee}, \citenamefont {Naylor}, \citenamefont {Ee},
  \citenamefont {Park}, \citenamefont {Johnson},\ and\ \citenamefont
  {Agarwal}}]{liu2016strong}%
  \BibitemOpen
  \bibfield  {author} {\bibinfo {author} {\bibfnamefont {W.}~\bibnamefont
  {Liu}}, \bibinfo {author} {\bibfnamefont {B.}~\bibnamefont {Lee}}, \bibinfo
  {author} {\bibfnamefont {C.~H.}\ \bibnamefont {Naylor}}, \bibinfo {author}
  {\bibfnamefont {H.-S.}\ \bibnamefont {Ee}}, \bibinfo {author} {\bibfnamefont
  {J.}~\bibnamefont {Park}}, \bibinfo {author} {\bibfnamefont {A.~C.}\
  \bibnamefont {Johnson}},\ and\ \bibinfo {author} {\bibfnamefont
  {R.}~\bibnamefont {Agarwal}},\ }\bibfield  {title} {\bibinfo {title} {Strong
  exciton--plasmon coupling in {M}o{S}$_{2}$ coupled with plasmonic lattice},\
  }\href@noop {} {\bibfield  {journal} {\bibinfo  {journal} {Nano Lett.}\
  }\textbf {\bibinfo {volume} {16}},\ \bibinfo {pages} {1262} (\bibinfo {year}
  {2016})}\BibitemShut {NoStop}%
\bibitem [{\citenamefont {Wang}\ \emph {et~al.}(2019)\citenamefont {Wang},
  \citenamefont {Le-Van}, \citenamefont {Vaianella}, \citenamefont {Maes},
  \citenamefont {Eizagirre~Barker}, \citenamefont {Godiksen}, \citenamefont
  {Curto},\ and\ \citenamefont {Gomez~Rivas}}]{wang2019limits}%
  \BibitemOpen
  \bibfield  {author} {\bibinfo {author} {\bibfnamefont {S.}~\bibnamefont
  {Wang}}, \bibinfo {author} {\bibfnamefont {Q.}~\bibnamefont {Le-Van}},
  \bibinfo {author} {\bibfnamefont {F.}~\bibnamefont {Vaianella}}, \bibinfo
  {author} {\bibfnamefont {B.}~\bibnamefont {Maes}}, \bibinfo {author}
  {\bibfnamefont {S.}~\bibnamefont {Eizagirre~Barker}}, \bibinfo {author}
  {\bibfnamefont {R.~H.}\ \bibnamefont {Godiksen}}, \bibinfo {author}
  {\bibfnamefont {A.~G.}\ \bibnamefont {Curto}},\ and\ \bibinfo {author}
  {\bibfnamefont {J.}~\bibnamefont {Gomez~Rivas}},\ }\bibfield  {title}
  {\bibinfo {title} {Limits to strong coupling of excitons in multilayer
  {WS}$_2$ with collective plasmonic resonances},\ }\href@noop {} {\bibfield
  {journal} {\bibinfo  {journal} {ACS Photonics}\ }\textbf {\bibinfo {volume}
  {6}},\ \bibinfo {pages} {286} (\bibinfo {year} {2019})}\BibitemShut {NoStop}%
\bibitem [{\citenamefont {Liu}\ \emph {et~al.}(2019)\citenamefont {Liu},
  \citenamefont {Wang}, \citenamefont {Zheng}, \citenamefont {Hwang},
  \citenamefont {Ji}, \citenamefont {Liu}, \citenamefont {Li}, \citenamefont
  {Sorger}, \citenamefont {Pan},\ and\ \citenamefont
  {Agarwal}}]{liu2019observation}%
  \BibitemOpen
  \bibfield  {author} {\bibinfo {author} {\bibfnamefont {W.}~\bibnamefont
  {Liu}}, \bibinfo {author} {\bibfnamefont {Y.}~\bibnamefont {Wang}}, \bibinfo
  {author} {\bibfnamefont {B.}~\bibnamefont {Zheng}}, \bibinfo {author}
  {\bibfnamefont {M.}~\bibnamefont {Hwang}}, \bibinfo {author} {\bibfnamefont
  {Z.}~\bibnamefont {Ji}}, \bibinfo {author} {\bibfnamefont {G.}~\bibnamefont
  {Liu}}, \bibinfo {author} {\bibfnamefont {Z.}~\bibnamefont {Li}}, \bibinfo
  {author} {\bibfnamefont {V.~J.}\ \bibnamefont {Sorger}}, \bibinfo {author}
  {\bibfnamefont {A.}~\bibnamefont {Pan}},\ and\ \bibinfo {author}
  {\bibfnamefont {R.}~\bibnamefont {Agarwal}},\ }\bibfield  {title} {\bibinfo
  {title} {Observation and active control of a collective polariton mode and
  polaritonic band gap in few-layer {WS}$_2$ strongly coupled with plasmonic
  lattices},\ }\href@noop {} {\bibfield  {journal} {\bibinfo  {journal} {Nano
  Lett.}\ }\textbf {\bibinfo {volume} {20}},\ \bibinfo {pages} {790} (\bibinfo
  {year} {2019})}\BibitemShut {NoStop}%
\bibitem [{\citenamefont {Khatibi}\ \emph {et~al.}(2018)\citenamefont
  {Khatibi}, \citenamefont {Feierabend}, \citenamefont {Selig}, \citenamefont
  {Brem}, \citenamefont {Linder{\"a}lv}, \citenamefont {Erhart},\ and\
  \citenamefont {Malic}}]{khatibi2018impact}%
  \BibitemOpen
  \bibfield  {author} {\bibinfo {author} {\bibfnamefont {Z.}~\bibnamefont
  {Khatibi}}, \bibinfo {author} {\bibfnamefont {M.}~\bibnamefont {Feierabend}},
  \bibinfo {author} {\bibfnamefont {M.}~\bibnamefont {Selig}}, \bibinfo
  {author} {\bibfnamefont {S.}~\bibnamefont {Brem}}, \bibinfo {author}
  {\bibfnamefont {C.}~\bibnamefont {Linder{\"a}lv}}, \bibinfo {author}
  {\bibfnamefont {P.}~\bibnamefont {Erhart}},\ and\ \bibinfo {author}
  {\bibfnamefont {E.}~\bibnamefont {Malic}},\ }\bibfield  {title} {\bibinfo
  {title} {Impact of strain on the excitonic linewidth in transition metal
  dichalcogenides},\ }\href@noop {} {\bibfield  {journal} {\bibinfo  {journal}
  {2D Mater.}\ }\textbf {\bibinfo {volume} {6}},\ \bibinfo {pages} {015015}
  (\bibinfo {year} {2018})}\BibitemShut {NoStop}%
\bibitem [{\citenamefont {Raja}\ \emph {et~al.}(2019)\citenamefont {Raja},
  \citenamefont {Waldecker}, \citenamefont {Zipfel}, \citenamefont {Cho},
  \citenamefont {Brem}, \citenamefont {Ziegler}, \citenamefont {Kulig},
  \citenamefont {Taniguchi}, \citenamefont {Watanabe}, \citenamefont {Malic},
  \citenamefont {Heinz}, \citenamefont {Berkelbach},\ and\ \citenamefont
  {Chernikov}}]{raja2019dielectric}%
  \BibitemOpen
  \bibfield  {author} {\bibinfo {author} {\bibfnamefont {A.}~\bibnamefont
  {Raja}}, \bibinfo {author} {\bibfnamefont {L.}~\bibnamefont {Waldecker}},
  \bibinfo {author} {\bibfnamefont {J.}~\bibnamefont {Zipfel}}, \bibinfo
  {author} {\bibfnamefont {Y.}~\bibnamefont {Cho}}, \bibinfo {author}
  {\bibfnamefont {S.}~\bibnamefont {Brem}}, \bibinfo {author} {\bibfnamefont
  {J.~D.}\ \bibnamefont {Ziegler}}, \bibinfo {author} {\bibfnamefont
  {M.}~\bibnamefont {Kulig}}, \bibinfo {author} {\bibfnamefont
  {T.}~\bibnamefont {Taniguchi}}, \bibinfo {author} {\bibfnamefont
  {K.}~\bibnamefont {Watanabe}}, \bibinfo {author} {\bibfnamefont
  {E.}~\bibnamefont {Malic}}, \bibinfo {author} {\bibfnamefont {T.~F.}\
  \bibnamefont {Heinz}}, \bibinfo {author} {\bibfnamefont {T.~C.}\ \bibnamefont
  {Berkelbach}},\ and\ \bibinfo {author} {\bibfnamefont {A.}~\bibnamefont
  {Chernikov}},\ }\bibfield  {title} {\bibinfo {title} {Dielectric disorder in
  two-dimensional materials},\ }\href@noop {} {\bibfield  {journal} {\bibinfo
  {journal} {Nat. Nanotechnol.}\ }\textbf {\bibinfo {volume} {14}},\ \bibinfo
  {pages} {832} (\bibinfo {year} {2019})}\BibitemShut {NoStop}%
\bibitem [{\citenamefont {Cadiz}\ \emph {et~al.}(2017)\citenamefont {Cadiz},
  \citenamefont {Courtade}, \citenamefont {Robert}, \citenamefont {Wang},
  \citenamefont {Shen}, \citenamefont {Cai}, \citenamefont {Taniguchi},
  \citenamefont {Watanabe}, \citenamefont {Carrere}, \citenamefont {Lagarde},
  \citenamefont {Manca}, \citenamefont {Amand}, \citenamefont {Renucci},
  \citenamefont {Tongay}, \citenamefont {Marie},\ and\ \citenamefont
  {Urbaszek}}]{cadiz2017excitonic}%
  \BibitemOpen
  \bibfield  {author} {\bibinfo {author} {\bibfnamefont {F.}~\bibnamefont
  {Cadiz}}, \bibinfo {author} {\bibfnamefont {E.}~\bibnamefont {Courtade}},
  \bibinfo {author} {\bibfnamefont {C.}~\bibnamefont {Robert}}, \bibinfo
  {author} {\bibfnamefont {G.}~\bibnamefont {Wang}}, \bibinfo {author}
  {\bibfnamefont {Y.}~\bibnamefont {Shen}}, \bibinfo {author} {\bibfnamefont
  {H.}~\bibnamefont {Cai}}, \bibinfo {author} {\bibfnamefont {T.}~\bibnamefont
  {Taniguchi}}, \bibinfo {author} {\bibfnamefont {K.}~\bibnamefont {Watanabe}},
  \bibinfo {author} {\bibfnamefont {H.}~\bibnamefont {Carrere}}, \bibinfo
  {author} {\bibfnamefont {D.}~\bibnamefont {Lagarde}}, \bibinfo {author}
  {\bibfnamefont {M.}~\bibnamefont {Manca}}, \bibinfo {author} {\bibfnamefont
  {T.}~\bibnamefont {Amand}}, \bibinfo {author} {\bibfnamefont
  {P.}~\bibnamefont {Renucci}}, \bibinfo {author} {\bibfnamefont
  {S.}~\bibnamefont {Tongay}}, \bibinfo {author} {\bibfnamefont
  {X.}~\bibnamefont {Marie}},\ and\ \bibinfo {author} {\bibfnamefont
  {B.}~\bibnamefont {Urbaszek}},\ }\bibfield  {title} {\bibinfo {title}
  {Excitonic linewidth approaching the homogeneous limit in {M}o{S}$_{2}$-based
  van der {W}aals heterostructures},\ }\href@noop {} {\bibfield  {journal}
  {\bibinfo  {journal} {Phys. Rev. X}\ }\textbf {\bibinfo {volume} {7}},\
  \bibinfo {pages} {021026} (\bibinfo {year} {2017})}\BibitemShut {NoStop}%
\bibitem [{\citenamefont {Wang}\ \emph
  {et~al.}(2016{\natexlab{a}})\citenamefont {Wang}, \citenamefont {Li},
  \citenamefont {Chervy}, \citenamefont {Shalabney}, \citenamefont {Azzini},
  \citenamefont {Orgiu}, \citenamefont {Hutchison}, \citenamefont {Genet},
  \citenamefont {Samor{\`\i}},\ and\ \citenamefont
  {Ebbesen}}]{wang2016coherent}%
  \BibitemOpen
  \bibfield  {author} {\bibinfo {author} {\bibfnamefont {S.}~\bibnamefont
  {Wang}}, \bibinfo {author} {\bibfnamefont {S.}~\bibnamefont {Li}}, \bibinfo
  {author} {\bibfnamefont {T.}~\bibnamefont {Chervy}}, \bibinfo {author}
  {\bibfnamefont {A.}~\bibnamefont {Shalabney}}, \bibinfo {author}
  {\bibfnamefont {S.}~\bibnamefont {Azzini}}, \bibinfo {author} {\bibfnamefont
  {E.}~\bibnamefont {Orgiu}}, \bibinfo {author} {\bibfnamefont {J.~A.}\
  \bibnamefont {Hutchison}}, \bibinfo {author} {\bibfnamefont {C.}~\bibnamefont
  {Genet}}, \bibinfo {author} {\bibfnamefont {P.}~\bibnamefont {Samor{\`\i}}},\
  and\ \bibinfo {author} {\bibfnamefont {T.~W.}\ \bibnamefont {Ebbesen}},\
  }\bibfield  {title} {\bibinfo {title} {Coherent coupling of {WS}$_2$
  monolayers with metallic photonic nanostructures at room temperature},\
  }\href@noop {} {\bibfield  {journal} {\bibinfo  {journal} {Nano Lett.}\
  }\textbf {\bibinfo {volume} {16}},\ \bibinfo {pages} {4368} (\bibinfo {year}
  {2016}{\natexlab{a}})}\BibitemShut {NoStop}%
\bibitem [{\citenamefont {Sun}\ \emph {et~al.}(2021)\citenamefont {Sun},
  \citenamefont {Li}, \citenamefont {Hu}, \citenamefont {Chen}, \citenamefont
  {Zheng}, \citenamefont {Zhang},\ and\ \citenamefont {Xu}}]{sun2021strong}%
  \BibitemOpen
  \bibfield  {author} {\bibinfo {author} {\bibfnamefont {J.}~\bibnamefont
  {Sun}}, \bibinfo {author} {\bibfnamefont {Y.}~\bibnamefont {Li}}, \bibinfo
  {author} {\bibfnamefont {H.}~\bibnamefont {Hu}}, \bibinfo {author}
  {\bibfnamefont {W.}~\bibnamefont {Chen}}, \bibinfo {author} {\bibfnamefont
  {D.}~\bibnamefont {Zheng}}, \bibinfo {author} {\bibfnamefont
  {S.}~\bibnamefont {Zhang}},\ and\ \bibinfo {author} {\bibfnamefont
  {H.}~\bibnamefont {Xu}},\ }\bibfield  {title} {\bibinfo {title} {Strong
  plasmon--exciton coupling in transition metal dichalcogenides and plasmonic
  nanostructures},\ }\href@noop {} {\bibfield  {journal} {\bibinfo  {journal}
  {Nanoscale}\ }\textbf {\bibinfo {volume} {13}},\ \bibinfo {pages} {4408}
  (\bibinfo {year} {2021})}\BibitemShut {NoStop}%
\bibitem [{\citenamefont {Guo}\ \emph {et~al.}(2023)\citenamefont {Guo},
  \citenamefont {Yu},\ and\ \citenamefont {Deng}}]{guo2023hybrid}%
  \BibitemOpen
  \bibfield  {author} {\bibinfo {author} {\bibfnamefont {C.}~\bibnamefont
  {Guo}}, \bibinfo {author} {\bibfnamefont {J.}~\bibnamefont {Yu}},\ and\
  \bibinfo {author} {\bibfnamefont {S.}~\bibnamefont {Deng}},\ }\bibfield
  {title} {\bibinfo {title} {Hybrid metasurfaces of plasmonic lattices and 2{D}
  materials},\ }\href@noop {} {\bibfield  {journal} {\bibinfo  {journal} {Adv.
  Funct. Mater.}\ }\textbf {\bibinfo {volume} {33}},\ \bibinfo {pages}
  {2302265} (\bibinfo {year} {2023})}\BibitemShut {NoStop}%
\bibitem [{\citenamefont {Klein}\ \emph {et~al.}(2019)\citenamefont {Klein},
  \citenamefont {Badada}, \citenamefont {Binder}, \citenamefont {Alfrey},
  \citenamefont {McKie}, \citenamefont {Koehler}, \citenamefont {Mandrus},
  \citenamefont {Taniguchi}, \citenamefont {Watanabe}, \citenamefont {LeRoy},\
  and\ \citenamefont {Schaibley}}]{klein20192d}%
  \BibitemOpen
  \bibfield  {author} {\bibinfo {author} {\bibfnamefont {M.}~\bibnamefont
  {Klein}}, \bibinfo {author} {\bibfnamefont {B.~H.}\ \bibnamefont {Badada}},
  \bibinfo {author} {\bibfnamefont {R.}~\bibnamefont {Binder}}, \bibinfo
  {author} {\bibfnamefont {A.}~\bibnamefont {Alfrey}}, \bibinfo {author}
  {\bibfnamefont {M.}~\bibnamefont {McKie}}, \bibinfo {author} {\bibfnamefont
  {M.~R.}\ \bibnamefont {Koehler}}, \bibinfo {author} {\bibfnamefont {D.~G.}\
  \bibnamefont {Mandrus}}, \bibinfo {author} {\bibfnamefont {T.}~\bibnamefont
  {Taniguchi}}, \bibinfo {author} {\bibfnamefont {K.}~\bibnamefont {Watanabe}},
  \bibinfo {author} {\bibfnamefont {B.~J.}\ \bibnamefont {LeRoy}},\ and\
  \bibinfo {author} {\bibfnamefont {J.~R.}\ \bibnamefont {Schaibley}},\
  }\bibfield  {title} {\bibinfo {title} {{2D} semiconductor nonlinear plasmonic
  modulators},\ }\href@noop {} {\bibfield  {journal} {\bibinfo  {journal} {Nat.
  Commun.}\ }\textbf {\bibinfo {volume} {10}},\ \bibinfo {pages} {3264}
  (\bibinfo {year} {2019})}\BibitemShut {NoStop}%
\bibitem [{\citenamefont {Dibos}\ \emph {et~al.}(2019)\citenamefont {Dibos},
  \citenamefont {Zhou}, \citenamefont {Jauregui}, \citenamefont {Scuri},
  \citenamefont {Wild}, \citenamefont {High}, \citenamefont {Taniguchi},
  \citenamefont {Watanabe}, \citenamefont {Lukin}, \citenamefont {Kim},\ and\
  \citenamefont {Park}}]{dibos2019electrically}%
  \BibitemOpen
  \bibfield  {author} {\bibinfo {author} {\bibfnamefont {A.~M.}\ \bibnamefont
  {Dibos}}, \bibinfo {author} {\bibfnamefont {Y.}~\bibnamefont {Zhou}},
  \bibinfo {author} {\bibfnamefont {L.~A.}\ \bibnamefont {Jauregui}}, \bibinfo
  {author} {\bibfnamefont {G.}~\bibnamefont {Scuri}}, \bibinfo {author}
  {\bibfnamefont {D.~S.}\ \bibnamefont {Wild}}, \bibinfo {author}
  {\bibfnamefont {A.~A.}\ \bibnamefont {High}}, \bibinfo {author}
  {\bibfnamefont {T.}~\bibnamefont {Taniguchi}}, \bibinfo {author}
  {\bibfnamefont {K.}~\bibnamefont {Watanabe}}, \bibinfo {author}
  {\bibfnamefont {M.~D.}\ \bibnamefont {Lukin}}, \bibinfo {author}
  {\bibfnamefont {P.}~\bibnamefont {Kim}},\ and\ \bibinfo {author}
  {\bibfnamefont {H.}~\bibnamefont {Park}},\ }\bibfield  {title} {\bibinfo
  {title} {Electrically tunable exciton--plasmon coupling in a {WSe}$_2$
  monolayer embedded in a plasmonic crystal cavity},\ }\href@noop {} {\bibfield
   {journal} {\bibinfo  {journal} {Nano Lett.}\ }\textbf {\bibinfo {volume}
  {19}},\ \bibinfo {pages} {3543} (\bibinfo {year} {2019})}\BibitemShut
  {NoStop}%
\bibitem [{\citenamefont {Vadia}\ \emph {et~al.}(2023)\citenamefont {Vadia},
  \citenamefont {Scherzer}, \citenamefont {Watanabe}, \citenamefont
  {Taniguchi},\ and\ \citenamefont {H{\"o}gele}}]{vadia2023magneto}%
  \BibitemOpen
  \bibfield  {author} {\bibinfo {author} {\bibfnamefont {S.}~\bibnamefont
  {Vadia}}, \bibinfo {author} {\bibfnamefont {J.}~\bibnamefont {Scherzer}},
  \bibinfo {author} {\bibfnamefont {K.}~\bibnamefont {Watanabe}}, \bibinfo
  {author} {\bibfnamefont {T.}~\bibnamefont {Taniguchi}},\ and\ \bibinfo
  {author} {\bibfnamefont {A.}~\bibnamefont {H{\"o}gele}},\ }\bibfield  {title}
  {\bibinfo {title} {Magneto-optical chirality in a coherently coupled
  exciton--plasmon system},\ }\href@noop {} {\bibfield  {journal} {\bibinfo
  {journal} {Nano Lett.}\ }\textbf {\bibinfo {volume} {23}},\ \bibinfo {pages}
  {614} (\bibinfo {year} {2023})}\BibitemShut {NoStop}%
\bibitem [{\citenamefont {Li}\ \emph {et~al.}(2022)\citenamefont {Li},
  \citenamefont {F{\"o}rste}, \citenamefont {Watanabe}, \citenamefont
  {Taniguchi}, \citenamefont {Urbaszek}, \citenamefont {Baimuratov},
  \citenamefont {Gerber}, \citenamefont {H{\"o}gele},\ and\ \citenamefont
  {Bilgin}}]{li2022stacking}%
  \BibitemOpen
  \bibfield  {author} {\bibinfo {author} {\bibfnamefont {Z.}~\bibnamefont
  {Li}}, \bibinfo {author} {\bibfnamefont {J.}~\bibnamefont {F{\"o}rste}},
  \bibinfo {author} {\bibfnamefont {K.}~\bibnamefont {Watanabe}}, \bibinfo
  {author} {\bibfnamefont {T.}~\bibnamefont {Taniguchi}}, \bibinfo {author}
  {\bibfnamefont {B.}~\bibnamefont {Urbaszek}}, \bibinfo {author}
  {\bibfnamefont {A.~S.}\ \bibnamefont {Baimuratov}}, \bibinfo {author}
  {\bibfnamefont {I.~C.}\ \bibnamefont {Gerber}}, \bibinfo {author}
  {\bibfnamefont {A.}~\bibnamefont {H{\"o}gele}},\ and\ \bibinfo {author}
  {\bibfnamefont {I.}~\bibnamefont {Bilgin}},\ }\bibfield  {title} {\bibinfo
  {title} {Stacking-dependent exciton multiplicity in {WSe}$_2$ bilayers},\
  }\href@noop {} {\bibfield  {journal} {\bibinfo  {journal} {Phys. Rev. B}\
  }\textbf {\bibinfo {volume} {106}},\ \bibinfo {pages} {045411} (\bibinfo
  {year} {2022})}\BibitemShut {NoStop}%
\bibitem [{\citenamefont {Baranov}\ \emph {et~al.}(2018)\citenamefont
  {Baranov}, \citenamefont {Wersall}, \citenamefont {Cuadra}, \citenamefont
  {Antosiewicz},\ and\ \citenamefont {Shegai}}]{baranov2018novel}%
  \BibitemOpen
  \bibfield  {author} {\bibinfo {author} {\bibfnamefont {D.~G.}\ \bibnamefont
  {Baranov}}, \bibinfo {author} {\bibfnamefont {M.}~\bibnamefont {Wersall}},
  \bibinfo {author} {\bibfnamefont {J.}~\bibnamefont {Cuadra}}, \bibinfo
  {author} {\bibfnamefont {T.~J.}\ \bibnamefont {Antosiewicz}},\ and\ \bibinfo
  {author} {\bibfnamefont {T.}~\bibnamefont {Shegai}},\ }\bibfield  {title}
  {\bibinfo {title} {Novel nanostructures and materials for strong
  light--matter interactions},\ }\href@noop {} {\bibfield  {journal} {\bibinfo
  {journal} {ACS Photonics}\ }\textbf {\bibinfo {volume} {5}},\ \bibinfo
  {pages} {24} (\bibinfo {year} {2018})}\BibitemShut {NoStop}%
\bibitem [{\citenamefont {Schneider}\ \emph {et~al.}(2018)\citenamefont
  {Schneider}, \citenamefont {Glazov}, \citenamefont {Korn}, \citenamefont
  {H{\"o}fling},\ and\ \citenamefont {Urbaszek}}]{schneider2018two}%
  \BibitemOpen
  \bibfield  {author} {\bibinfo {author} {\bibfnamefont {C.}~\bibnamefont
  {Schneider}}, \bibinfo {author} {\bibfnamefont {M.~M.}\ \bibnamefont
  {Glazov}}, \bibinfo {author} {\bibfnamefont {T.}~\bibnamefont {Korn}},
  \bibinfo {author} {\bibfnamefont {S.}~\bibnamefont {H{\"o}fling}},\ and\
  \bibinfo {author} {\bibfnamefont {B.}~\bibnamefont {Urbaszek}},\ }\bibfield
  {title} {\bibinfo {title} {Two-dimensional semiconductors in the regime of
  strong light-matter coupling},\ }\href@noop {} {\bibfield  {journal}
  {\bibinfo  {journal} {Nat. Commun.}\ }\textbf {\bibinfo {volume} {9}},\
  \bibinfo {pages} {2695} (\bibinfo {year} {2018})}\BibitemShut {NoStop}%
\bibitem [{\citenamefont {You}\ \emph {et~al.}(2015)\citenamefont {You},
  \citenamefont {Zhang}, \citenamefont {Berkelbach}, \citenamefont {Hybertsen},
  \citenamefont {Reichman},\ and\ \citenamefont {Heinz}}]{you2015observation}%
  \BibitemOpen
  \bibfield  {author} {\bibinfo {author} {\bibfnamefont {Y.}~\bibnamefont
  {You}}, \bibinfo {author} {\bibfnamefont {X.-X.}\ \bibnamefont {Zhang}},
  \bibinfo {author} {\bibfnamefont {T.~C.}\ \bibnamefont {Berkelbach}},
  \bibinfo {author} {\bibfnamefont {M.~S.}\ \bibnamefont {Hybertsen}}, \bibinfo
  {author} {\bibfnamefont {D.~R.}\ \bibnamefont {Reichman}},\ and\ \bibinfo
  {author} {\bibfnamefont {T.~F.}\ \bibnamefont {Heinz}},\ }\bibfield  {title}
  {\bibinfo {title} {Observation of biexcitons in monolayer {WSe}$_2$},\
  }\href@noop {} {\bibfield  {journal} {\bibinfo  {journal} {Nat. Phys.}\
  }\textbf {\bibinfo {volume} {11}},\ \bibinfo {pages} {477} (\bibinfo {year}
  {2015})}\BibitemShut {NoStop}%
\bibitem [{\citenamefont {Barbone}\ \emph {et~al.}(2018)\citenamefont
  {Barbone}, \citenamefont {Montblanch}, \citenamefont {Kara}, \citenamefont
  {Palacios-Berraquero}, \citenamefont {Cadore}, \citenamefont {De~Fazio},
  \citenamefont {Pingault}, \citenamefont {Mostaani}, \citenamefont {Li},
  \citenamefont {Chen}, \citenamefont {Watanabe}, \citenamefont {Taniguchi},
  \citenamefont {Tongay}, \citenamefont {amd Andrea C.~Ferrari},\ and\
  \citenamefont {Atat{\"u}re}}]{barbone2018charge}%
  \BibitemOpen
  \bibfield  {author} {\bibinfo {author} {\bibfnamefont {M.}~\bibnamefont
  {Barbone}}, \bibinfo {author} {\bibfnamefont {A.~R.-P.}\ \bibnamefont
  {Montblanch}}, \bibinfo {author} {\bibfnamefont {D.~M.}\ \bibnamefont
  {Kara}}, \bibinfo {author} {\bibfnamefont {C.}~\bibnamefont
  {Palacios-Berraquero}}, \bibinfo {author} {\bibfnamefont {A.~R.}\
  \bibnamefont {Cadore}}, \bibinfo {author} {\bibfnamefont {D.}~\bibnamefont
  {De~Fazio}}, \bibinfo {author} {\bibfnamefont {B.}~\bibnamefont {Pingault}},
  \bibinfo {author} {\bibfnamefont {E.}~\bibnamefont {Mostaani}}, \bibinfo
  {author} {\bibfnamefont {H.}~\bibnamefont {Li}}, \bibinfo {author}
  {\bibfnamefont {B.}~\bibnamefont {Chen}}, \bibinfo {author} {\bibfnamefont
  {K.}~\bibnamefont {Watanabe}}, \bibinfo {author} {\bibfnamefont
  {T.}~\bibnamefont {Taniguchi}}, \bibinfo {author} {\bibfnamefont
  {S.}~\bibnamefont {Tongay}}, \bibinfo {author} {\bibfnamefont {G.~W.}\
  \bibnamefont {amd Andrea C.~Ferrari}},\ and\ \bibinfo {author} {\bibfnamefont
  {M.}~\bibnamefont {Atat{\"u}re}},\ }\bibfield  {title} {\bibinfo {title}
  {Charge-tuneable biexciton complexes in monolayer {WSe}$_2$},\ }\href@noop {}
  {\bibfield  {journal} {\bibinfo  {journal} {Nat. Commun.}\ }\textbf {\bibinfo
  {volume} {9}},\ \bibinfo {pages} {3721} (\bibinfo {year} {2018})}\BibitemShut
  {NoStop}%
\bibitem [{\citenamefont {Li}\ \emph {et~al.}(2018)\citenamefont {Li},
  \citenamefont {Wang}, \citenamefont {Lu}, \citenamefont {Jin}, \citenamefont
  {Chen}, \citenamefont {Meng}, \citenamefont {Lian}, \citenamefont
  {Taniguchi}, \citenamefont {Watanabe}, \citenamefont {Zhang}, \citenamefont
  {Smirnov},\ and\ \citenamefont {Shi}}]{li2018revealing}%
  \BibitemOpen
  \bibfield  {author} {\bibinfo {author} {\bibfnamefont {Z.}~\bibnamefont
  {Li}}, \bibinfo {author} {\bibfnamefont {T.}~\bibnamefont {Wang}}, \bibinfo
  {author} {\bibfnamefont {Z.}~\bibnamefont {Lu}}, \bibinfo {author}
  {\bibfnamefont {C.}~\bibnamefont {Jin}}, \bibinfo {author} {\bibfnamefont
  {Y.}~\bibnamefont {Chen}}, \bibinfo {author} {\bibfnamefont {Y.}~\bibnamefont
  {Meng}}, \bibinfo {author} {\bibfnamefont {Z.}~\bibnamefont {Lian}}, \bibinfo
  {author} {\bibfnamefont {T.}~\bibnamefont {Taniguchi}}, \bibinfo {author}
  {\bibfnamefont {K.}~\bibnamefont {Watanabe}}, \bibinfo {author}
  {\bibfnamefont {S.}~\bibnamefont {Zhang}}, \bibinfo {author} {\bibfnamefont
  {D.}~\bibnamefont {Smirnov}},\ and\ \bibinfo {author} {\bibfnamefont {S.-F.}\
  \bibnamefont {Shi}},\ }\bibfield  {title} {\bibinfo {title} {Revealing the
  biexciton and trion-exciton complexes in {BN} encapsulated {WSe}$_2$},\
  }\href@noop {} {\bibfield  {journal} {\bibinfo  {journal} {Nat. Commun.}\
  }\textbf {\bibinfo {volume} {9}},\ \bibinfo {pages} {3719} (\bibinfo {year}
  {2018})}\BibitemShut {NoStop}%
\bibitem [{\citenamefont {Steinhoff}\ \emph {et~al.}(2018)\citenamefont
  {Steinhoff}, \citenamefont {Florian}, \citenamefont {Singh}, \citenamefont
  {Tran}, \citenamefont {Kolarczik}, \citenamefont {Helmrich}, \citenamefont
  {Achtstein}, \citenamefont {Woggon}, \citenamefont {Owschimikow},
  \citenamefont {Jahnke},\ and\ \citenamefont {Li}}]{steinhoff2018biexciton}%
  \BibitemOpen
  \bibfield  {author} {\bibinfo {author} {\bibfnamefont {A.}~\bibnamefont
  {Steinhoff}}, \bibinfo {author} {\bibfnamefont {M.}~\bibnamefont {Florian}},
  \bibinfo {author} {\bibfnamefont {A.}~\bibnamefont {Singh}}, \bibinfo
  {author} {\bibfnamefont {K.}~\bibnamefont {Tran}}, \bibinfo {author}
  {\bibfnamefont {M.}~\bibnamefont {Kolarczik}}, \bibinfo {author}
  {\bibfnamefont {S.}~\bibnamefont {Helmrich}}, \bibinfo {author}
  {\bibfnamefont {A.~W.}\ \bibnamefont {Achtstein}}, \bibinfo {author}
  {\bibfnamefont {U.}~\bibnamefont {Woggon}}, \bibinfo {author} {\bibfnamefont
  {N.}~\bibnamefont {Owschimikow}}, \bibinfo {author} {\bibfnamefont
  {F.}~\bibnamefont {Jahnke}},\ and\ \bibinfo {author} {\bibfnamefont
  {X.}~\bibnamefont {Li}},\ }\bibfield  {title} {\bibinfo {title} {Biexciton
  fine structure in monolayer transition metal dichalcogenides},\ }\href@noop
  {} {\bibfield  {journal} {\bibinfo  {journal} {Nat. Phys.}\ }\textbf
  {\bibinfo {volume} {14}},\ \bibinfo {pages} {1199} (\bibinfo {year}
  {2018})}\BibitemShut {NoStop}%
\bibitem [{\citenamefont {Zhang}\ \emph {et~al.}(2015)\citenamefont {Zhang},
  \citenamefont {You}, \citenamefont {Zhao},\ and\ \citenamefont
  {Heinz}}]{zhang2015experimental}%
  \BibitemOpen
  \bibfield  {author} {\bibinfo {author} {\bibfnamefont {X.-X.}\ \bibnamefont
  {Zhang}}, \bibinfo {author} {\bibfnamefont {Y.}~\bibnamefont {You}}, \bibinfo
  {author} {\bibfnamefont {S.~Y.~F.}\ \bibnamefont {Zhao}},\ and\ \bibinfo
  {author} {\bibfnamefont {T.~F.}\ \bibnamefont {Heinz}},\ }\bibfield  {title}
  {\bibinfo {title} {Experimental evidence for dark excitons in monolayer
  {WSe}$_2$},\ }\href@noop {} {\bibfield  {journal} {\bibinfo  {journal} {Phys.
  Rev. Lett.}\ }\textbf {\bibinfo {volume} {115}},\ \bibinfo {pages} {257403}
  (\bibinfo {year} {2015})}\BibitemShut {NoStop}%
\bibitem [{\citenamefont {Robert}\ \emph {et~al.}(2017)\citenamefont {Robert},
  \citenamefont {Amand}, \citenamefont {Cadiz}, \citenamefont {Lagarde},
  \citenamefont {Courtade}, \citenamefont {Manca}, \citenamefont {Taniguchi},
  \citenamefont {Watanabe}, \citenamefont {Urbaszek},\ and\ \citenamefont
  {Marie}}]{robert2017fine}%
  \BibitemOpen
  \bibfield  {author} {\bibinfo {author} {\bibfnamefont {C.}~\bibnamefont
  {Robert}}, \bibinfo {author} {\bibfnamefont {T.}~\bibnamefont {Amand}},
  \bibinfo {author} {\bibfnamefont {F.}~\bibnamefont {Cadiz}}, \bibinfo
  {author} {\bibfnamefont {D.}~\bibnamefont {Lagarde}}, \bibinfo {author}
  {\bibfnamefont {E.}~\bibnamefont {Courtade}}, \bibinfo {author}
  {\bibfnamefont {M.}~\bibnamefont {Manca}}, \bibinfo {author} {\bibfnamefont
  {T.}~\bibnamefont {Taniguchi}}, \bibinfo {author} {\bibfnamefont
  {K.}~\bibnamefont {Watanabe}}, \bibinfo {author} {\bibfnamefont
  {B.}~\bibnamefont {Urbaszek}},\ and\ \bibinfo {author} {\bibfnamefont
  {X.}~\bibnamefont {Marie}},\ }\bibfield  {title} {\bibinfo {title} {Fine
  structure and lifetime of dark excitons in transition metal dichalcogenide
  monolayers},\ }\href@noop {} {\bibfield  {journal} {\bibinfo  {journal}
  {Phys. Rev. B}\ }\textbf {\bibinfo {volume} {96}},\ \bibinfo {pages} {155423}
  (\bibinfo {year} {2017})}\BibitemShut {NoStop}%
\bibitem [{\citenamefont {Courtade}\ \emph {et~al.}(2017)\citenamefont
  {Courtade}, \citenamefont {Semina}, \citenamefont {Manca}, \citenamefont
  {Glazov}, \citenamefont {Robert}, \citenamefont {Cadiz}, \citenamefont
  {Wang}, \citenamefont {Taniguchi}, \citenamefont {Watanabe}, \citenamefont
  {Pierre}, \citenamefont {Escoffier}, \citenamefont {Ivchenko}, \citenamefont
  {Renucci}, \citenamefont {Marie}, \citenamefont {Amand},\ and\ \citenamefont
  {Urbaszek}}]{courtade2017charged}%
  \BibitemOpen
  \bibfield  {author} {\bibinfo {author} {\bibfnamefont {E.}~\bibnamefont
  {Courtade}}, \bibinfo {author} {\bibfnamefont {M.}~\bibnamefont {Semina}},
  \bibinfo {author} {\bibfnamefont {M.}~\bibnamefont {Manca}}, \bibinfo
  {author} {\bibfnamefont {M.}~\bibnamefont {Glazov}}, \bibinfo {author}
  {\bibfnamefont {C.}~\bibnamefont {Robert}}, \bibinfo {author} {\bibfnamefont
  {F.}~\bibnamefont {Cadiz}}, \bibinfo {author} {\bibfnamefont
  {G.}~\bibnamefont {Wang}}, \bibinfo {author} {\bibfnamefont {T.}~\bibnamefont
  {Taniguchi}}, \bibinfo {author} {\bibfnamefont {K.}~\bibnamefont {Watanabe}},
  \bibinfo {author} {\bibfnamefont {M.}~\bibnamefont {Pierre}}, \bibinfo
  {author} {\bibfnamefont {W.}~\bibnamefont {Escoffier}}, \bibinfo {author}
  {\bibfnamefont {E.~L.}\ \bibnamefont {Ivchenko}}, \bibinfo {author}
  {\bibfnamefont {P.}~\bibnamefont {Renucci}}, \bibinfo {author} {\bibfnamefont
  {X.}~\bibnamefont {Marie}}, \bibinfo {author} {\bibfnamefont
  {T.}~\bibnamefont {Amand}},\ and\ \bibinfo {author} {\bibfnamefont
  {B.}~\bibnamefont {Urbaszek}},\ }\bibfield  {title} {\bibinfo {title}
  {Charged excitons in monolayer {WSe}$_2$: Experiment and theory},\
  }\href@noop {} {\bibfield  {journal} {\bibinfo  {journal} {Phys. Rev. B}\
  }\textbf {\bibinfo {volume} {96}},\ \bibinfo {pages} {085302} (\bibinfo
  {year} {2017})}\BibitemShut {NoStop}%
\bibitem [{\citenamefont {He}\ \emph {et~al.}(2020)\citenamefont {He},
  \citenamefont {Rivera}, \citenamefont {Van~Tuan}, \citenamefont {Wilson},
  \citenamefont {Yang}, \citenamefont {Taniguchi}, \citenamefont {Watanabe},
  \citenamefont {Yan}, \citenamefont {Mandrus}, \citenamefont {Yu},
  \citenamefont {Dery}, \citenamefont {Yao},\ and\ \citenamefont
  {Xu}}]{he2020valley}%
  \BibitemOpen
  \bibfield  {author} {\bibinfo {author} {\bibfnamefont {M.}~\bibnamefont
  {He}}, \bibinfo {author} {\bibfnamefont {P.}~\bibnamefont {Rivera}}, \bibinfo
  {author} {\bibfnamefont {D.}~\bibnamefont {Van~Tuan}}, \bibinfo {author}
  {\bibfnamefont {N.~P.}\ \bibnamefont {Wilson}}, \bibinfo {author}
  {\bibfnamefont {M.}~\bibnamefont {Yang}}, \bibinfo {author} {\bibfnamefont
  {T.}~\bibnamefont {Taniguchi}}, \bibinfo {author} {\bibfnamefont
  {K.}~\bibnamefont {Watanabe}}, \bibinfo {author} {\bibfnamefont
  {J.}~\bibnamefont {Yan}}, \bibinfo {author} {\bibfnamefont {D.~G.}\
  \bibnamefont {Mandrus}}, \bibinfo {author} {\bibfnamefont {H.}~\bibnamefont
  {Yu}}, \bibinfo {author} {\bibfnamefont {H.}~\bibnamefont {Dery}}, \bibinfo
  {author} {\bibfnamefont {W.}~\bibnamefont {Yao}},\ and\ \bibinfo {author}
  {\bibfnamefont {X.}~\bibnamefont {Xu}},\ }\bibfield  {title} {\bibinfo
  {title} {Valley phonons and exciton complexes in a monolayer semiconductor},\
  }\href@noop {} {\bibfield  {journal} {\bibinfo  {journal} {Nat. Commun.}\
  }\textbf {\bibinfo {volume} {11}},\ \bibinfo {pages} {618} (\bibinfo {year}
  {2020})}\BibitemShut {NoStop}%
\bibitem [{\citenamefont {Liu}\ \emph {et~al.}(2020)\citenamefont {Liu},
  \citenamefont {van Baren}, \citenamefont {Liang}, \citenamefont {Taniguchi},
  \citenamefont {Watanabe}, \citenamefont {Gabor}, \citenamefont {Chang},\ and\
  \citenamefont {Lui}}]{liu2020multipath}%
  \BibitemOpen
  \bibfield  {author} {\bibinfo {author} {\bibfnamefont {E.}~\bibnamefont
  {Liu}}, \bibinfo {author} {\bibfnamefont {J.}~\bibnamefont {van Baren}},
  \bibinfo {author} {\bibfnamefont {C.-T.}\ \bibnamefont {Liang}}, \bibinfo
  {author} {\bibfnamefont {T.}~\bibnamefont {Taniguchi}}, \bibinfo {author}
  {\bibfnamefont {K.}~\bibnamefont {Watanabe}}, \bibinfo {author}
  {\bibfnamefont {N.~M.}\ \bibnamefont {Gabor}}, \bibinfo {author}
  {\bibfnamefont {Y.-C.}\ \bibnamefont {Chang}},\ and\ \bibinfo {author}
  {\bibfnamefont {C.~H.}\ \bibnamefont {Lui}},\ }\bibfield  {title} {\bibinfo
  {title} {Multipath optical recombination of intervalley dark excitons and
  trions in monolayer {WSe}$_2$},\ }\href@noop {} {\bibfield  {journal}
  {\bibinfo  {journal} {Phys. Rev. Lett.}\ }\textbf {\bibinfo {volume} {124}},\
  \bibinfo {pages} {196802} (\bibinfo {year} {2020})}\BibitemShut {NoStop}%
\bibitem [{\citenamefont {Yang}\ \emph {et~al.}(2022)\citenamefont {Yang},
  \citenamefont {Ren}, \citenamefont {Robert}, \citenamefont {Van~Tuan},
  \citenamefont {Lombez}, \citenamefont {Urbaszek}, \citenamefont {Marie},\
  and\ \citenamefont {Dery}}]{yang2022relaxation}%
  \BibitemOpen
  \bibfield  {author} {\bibinfo {author} {\bibfnamefont {M.}~\bibnamefont
  {Yang}}, \bibinfo {author} {\bibfnamefont {L.}~\bibnamefont {Ren}}, \bibinfo
  {author} {\bibfnamefont {C.}~\bibnamefont {Robert}}, \bibinfo {author}
  {\bibfnamefont {D.}~\bibnamefont {Van~Tuan}}, \bibinfo {author}
  {\bibfnamefont {L.}~\bibnamefont {Lombez}}, \bibinfo {author} {\bibfnamefont
  {B.}~\bibnamefont {Urbaszek}}, \bibinfo {author} {\bibfnamefont
  {X.}~\bibnamefont {Marie}},\ and\ \bibinfo {author} {\bibfnamefont
  {H.}~\bibnamefont {Dery}},\ }\bibfield  {title} {\bibinfo {title} {Relaxation
  and darkening of excitonic complexes in electrostatically doped monolayer
  {WSe}$_2$: Roles of exciton-electron and trion-electron interactions},\
  }\href@noop {} {\bibfield  {journal} {\bibinfo  {journal} {Phys. Rev. B}\
  }\textbf {\bibinfo {volume} {105}},\ \bibinfo {pages} {085302} (\bibinfo
  {year} {2022})}\BibitemShut {NoStop}%
\bibitem [{\citenamefont {F{\"o}rste}\ \emph {et~al.}(2020)\citenamefont
  {F{\"o}rste}, \citenamefont {Tepliakov}, \citenamefont {Kruchinin},
  \citenamefont {Lindlau}, \citenamefont {Funk}, \citenamefont {F{\"o}rg},
  \citenamefont {Watanabe}, \citenamefont {Taniguchi}, \citenamefont
  {Baimuratov},\ and\ \citenamefont {H{\"o}gele}}]{forste2020exciton}%
  \BibitemOpen
  \bibfield  {author} {\bibinfo {author} {\bibfnamefont {J.}~\bibnamefont
  {F{\"o}rste}}, \bibinfo {author} {\bibfnamefont {N.~V.}\ \bibnamefont
  {Tepliakov}}, \bibinfo {author} {\bibfnamefont {S.~Y.}\ \bibnamefont
  {Kruchinin}}, \bibinfo {author} {\bibfnamefont {J.}~\bibnamefont {Lindlau}},
  \bibinfo {author} {\bibfnamefont {V.}~\bibnamefont {Funk}}, \bibinfo {author}
  {\bibfnamefont {M.}~\bibnamefont {F{\"o}rg}}, \bibinfo {author}
  {\bibfnamefont {K.}~\bibnamefont {Watanabe}}, \bibinfo {author}
  {\bibfnamefont {T.}~\bibnamefont {Taniguchi}}, \bibinfo {author}
  {\bibfnamefont {A.~S.}\ \bibnamefont {Baimuratov}},\ and\ \bibinfo {author}
  {\bibfnamefont {A.}~\bibnamefont {H{\"o}gele}},\ }\bibfield  {title}
  {\bibinfo {title} {Exciton g-factors in monolayer and bilayer {WSe}$_2$ from
  experiment and theory},\ }\href@noop {} {\bibfield  {journal} {\bibinfo
  {journal} {Nat. Commun.}\ }\textbf {\bibinfo {volume} {11}},\ \bibinfo
  {pages} {4539} (\bibinfo {year} {2020})}\BibitemShut {NoStop}%
\bibitem [{\citenamefont {Zeng}\ \emph {et~al.}(2012)\citenamefont {Zeng},
  \citenamefont {Dai}, \citenamefont {Yao}, \citenamefont {Xiao},\ and\
  \citenamefont {Cui}}]{zeng2012valley}%
  \BibitemOpen
  \bibfield  {author} {\bibinfo {author} {\bibfnamefont {H.}~\bibnamefont
  {Zeng}}, \bibinfo {author} {\bibfnamefont {J.}~\bibnamefont {Dai}}, \bibinfo
  {author} {\bibfnamefont {W.}~\bibnamefont {Yao}}, \bibinfo {author}
  {\bibfnamefont {D.}~\bibnamefont {Xiao}},\ and\ \bibinfo {author}
  {\bibfnamefont {X.}~\bibnamefont {Cui}},\ }\bibfield  {title} {\bibinfo
  {title} {Valley polarization in {MoS}$_2$ monolayers by optical pumping},\
  }\href@noop {} {\bibfield  {journal} {\bibinfo  {journal} {Nat.
  Nanotechnol.}\ }\textbf {\bibinfo {volume} {7}},\ \bibinfo {pages} {490}
  (\bibinfo {year} {2012})}\BibitemShut {NoStop}%
\bibitem [{\citenamefont {Jones}\ \emph {et~al.}(2013)\citenamefont {Jones},
  \citenamefont {Yu}, \citenamefont {Ghimire}, \citenamefont {Wu},
  \citenamefont {Aivazian}, \citenamefont {Ross}, \citenamefont {Zhao},
  \citenamefont {Yan}, \citenamefont {Mandrus}, \citenamefont {Xiao},
  \citenamefont {Yao},\ and\ \citenamefont {Xu}}]{jones2013optical}%
  \BibitemOpen
  \bibfield  {author} {\bibinfo {author} {\bibfnamefont {A.~M.}\ \bibnamefont
  {Jones}}, \bibinfo {author} {\bibfnamefont {H.}~\bibnamefont {Yu}}, \bibinfo
  {author} {\bibfnamefont {N.~J.}\ \bibnamefont {Ghimire}}, \bibinfo {author}
  {\bibfnamefont {S.}~\bibnamefont {Wu}}, \bibinfo {author} {\bibfnamefont
  {G.}~\bibnamefont {Aivazian}}, \bibinfo {author} {\bibfnamefont {J.~S.}\
  \bibnamefont {Ross}}, \bibinfo {author} {\bibfnamefont {B.}~\bibnamefont
  {Zhao}}, \bibinfo {author} {\bibfnamefont {J.}~\bibnamefont {Yan}}, \bibinfo
  {author} {\bibfnamefont {D.~G.}\ \bibnamefont {Mandrus}}, \bibinfo {author}
  {\bibfnamefont {D.}~\bibnamefont {Xiao}}, \bibinfo {author} {\bibfnamefont
  {W.}~\bibnamefont {Yao}},\ and\ \bibinfo {author} {\bibfnamefont
  {X.}~\bibnamefont {Xu}},\ }\bibfield  {title} {\bibinfo {title} {Optical
  generation of excitonic valley coherence in monolayer {WSe}$_2$},\
  }\href@noop {} {\bibfield  {journal} {\bibinfo  {journal} {Nat.
  Nanotechnol.}\ }\textbf {\bibinfo {volume} {8}},\ \bibinfo {pages} {634}
  (\bibinfo {year} {2013})}\BibitemShut {NoStop}%
\bibitem [{\citenamefont {Wang}\ \emph
  {et~al.}(2016{\natexlab{b}})\citenamefont {Wang}, \citenamefont {Marie},
  \citenamefont {Liu}, \citenamefont {Amand}, \citenamefont {Robert},
  \citenamefont {Cadiz}, \citenamefont {Renucci},\ and\ \citenamefont
  {Urbaszek}}]{wang2016control}%
  \BibitemOpen
  \bibfield  {author} {\bibinfo {author} {\bibfnamefont {G.}~\bibnamefont
  {Wang}}, \bibinfo {author} {\bibfnamefont {X.}~\bibnamefont {Marie}},
  \bibinfo {author} {\bibfnamefont {B.}~\bibnamefont {Liu}}, \bibinfo {author}
  {\bibfnamefont {T.}~\bibnamefont {Amand}}, \bibinfo {author} {\bibfnamefont
  {C.}~\bibnamefont {Robert}}, \bibinfo {author} {\bibfnamefont
  {F.}~\bibnamefont {Cadiz}}, \bibinfo {author} {\bibfnamefont
  {P.}~\bibnamefont {Renucci}},\ and\ \bibinfo {author} {\bibfnamefont
  {B.}~\bibnamefont {Urbaszek}},\ }\bibfield  {title} {\bibinfo {title}
  {Control of exciton valley coherence in transition metal dichalcogenide
  monolayers},\ }\href@noop {} {\bibfield  {journal} {\bibinfo  {journal}
  {Phys. Rev. Lett.}\ }\textbf {\bibinfo {volume} {117}},\ \bibinfo {pages}
  {187401} (\bibinfo {year} {2016}{\natexlab{b}})}\BibitemShut {NoStop}%
\bibitem [{\citenamefont {Gong}\ \emph {et~al.}(2018)\citenamefont {Gong},
  \citenamefont {Alpeggiani}, \citenamefont {Sciacca}, \citenamefont
  {Garnett},\ and\ \citenamefont {Kuipers}}]{gong2018nanoscale}%
  \BibitemOpen
  \bibfield  {author} {\bibinfo {author} {\bibfnamefont {S.-H.}\ \bibnamefont
  {Gong}}, \bibinfo {author} {\bibfnamefont {F.}~\bibnamefont {Alpeggiani}},
  \bibinfo {author} {\bibfnamefont {B.}~\bibnamefont {Sciacca}}, \bibinfo
  {author} {\bibfnamefont {E.~C.}\ \bibnamefont {Garnett}},\ and\ \bibinfo
  {author} {\bibfnamefont {L.}~\bibnamefont {Kuipers}},\ }\bibfield  {title}
  {\bibinfo {title} {Nanoscale chiral valley-photon interface through optical
  spin-orbit coupling},\ }\href@noop {} {\bibfield  {journal} {\bibinfo
  {journal} {Science}\ }\textbf {\bibinfo {volume} {359}},\ \bibinfo {pages}
  {443} (\bibinfo {year} {2018})}\BibitemShut {NoStop}%
\bibitem [{\citenamefont {Hu}\ \emph {et~al.}(2019)\citenamefont {Hu},
  \citenamefont {Hong}, \citenamefont {Wang}, \citenamefont {Wu}, \citenamefont
  {Xu}, \citenamefont {Zhao}, \citenamefont {Liu}, \citenamefont {Zhang},
  \citenamefont {Garcia-Vidal}, \citenamefont {Wang}, \citenamefont {Lu},\ and\
  \citenamefont {Qiu}}]{hu2019coherent}%
  \BibitemOpen
  \bibfield  {author} {\bibinfo {author} {\bibfnamefont {G.}~\bibnamefont
  {Hu}}, \bibinfo {author} {\bibfnamefont {X.}~\bibnamefont {Hong}}, \bibinfo
  {author} {\bibfnamefont {K.}~\bibnamefont {Wang}}, \bibinfo {author}
  {\bibfnamefont {J.}~\bibnamefont {Wu}}, \bibinfo {author} {\bibfnamefont
  {H.-X.}\ \bibnamefont {Xu}}, \bibinfo {author} {\bibfnamefont
  {W.}~\bibnamefont {Zhao}}, \bibinfo {author} {\bibfnamefont {W.}~\bibnamefont
  {Liu}}, \bibinfo {author} {\bibfnamefont {S.}~\bibnamefont {Zhang}}, \bibinfo
  {author} {\bibfnamefont {F.}~\bibnamefont {Garcia-Vidal}}, \bibinfo {author}
  {\bibfnamefont {B.}~\bibnamefont {Wang}}, \bibinfo {author} {\bibfnamefont
  {P.}~\bibnamefont {Lu}},\ and\ \bibinfo {author} {\bibfnamefont {C.-W.}\
  \bibnamefont {Qiu}},\ }\bibfield  {title} {\bibinfo {title} {Coherent
  steering of nonlinear chiral valley photons with a synthetic {Au}--{WS}$_2$
  metasurface},\ }\href@noop {} {\bibfield  {journal} {\bibinfo  {journal}
  {Nat. Photonics}\ }\textbf {\bibinfo {volume} {13}},\ \bibinfo {pages} {467}
  (\bibinfo {year} {2019})}\BibitemShut {NoStop}%
\bibitem [{\citenamefont {Guddala}\ \emph {et~al.}(2019)\citenamefont
  {Guddala}, \citenamefont {Bushati}, \citenamefont {Li}, \citenamefont
  {Khanikaev},\ and\ \citenamefont {Menon}}]{guddala2019valley}%
  \BibitemOpen
  \bibfield  {author} {\bibinfo {author} {\bibfnamefont {S.}~\bibnamefont
  {Guddala}}, \bibinfo {author} {\bibfnamefont {R.}~\bibnamefont {Bushati}},
  \bibinfo {author} {\bibfnamefont {M.}~\bibnamefont {Li}}, \bibinfo {author}
  {\bibfnamefont {A.}~\bibnamefont {Khanikaev}},\ and\ \bibinfo {author}
  {\bibfnamefont {V.}~\bibnamefont {Menon}},\ }\bibfield  {title} {\bibinfo
  {title} {Valley selective optical control of excitons in {2D} semiconductors
  using a chiral metasurface},\ }\href@noop {} {\bibfield  {journal} {\bibinfo
  {journal} {Opt. Mater. Express}\ }\textbf {\bibinfo {volume} {9}},\ \bibinfo
  {pages} {536} (\bibinfo {year} {2019})}\BibitemShut {NoStop}%
\bibitem [{\citenamefont {Ozawa}\ \emph {et~al.}(2019)\citenamefont {Ozawa},
  \citenamefont {Price}, \citenamefont {Amo}, \citenamefont {Goldman},
  \citenamefont {Hafezi}, \citenamefont {Lu}, \citenamefont {Rechtsman},
  \citenamefont {Schuster}, \citenamefont {Simon}, \citenamefont {Zilberberg},\
  and\ \citenamefont {Carusotto}}]{Ozawa2019}%
  \BibitemOpen
  \bibfield  {author} {\bibinfo {author} {\bibfnamefont {T.}~\bibnamefont
  {Ozawa}}, \bibinfo {author} {\bibfnamefont {H.~M.}\ \bibnamefont {Price}},
  \bibinfo {author} {\bibfnamefont {A.}~\bibnamefont {Amo}}, \bibinfo {author}
  {\bibfnamefont {N.}~\bibnamefont {Goldman}}, \bibinfo {author} {\bibfnamefont
  {M.}~\bibnamefont {Hafezi}}, \bibinfo {author} {\bibfnamefont
  {L.}~\bibnamefont {Lu}}, \bibinfo {author} {\bibfnamefont {M.~C.}\
  \bibnamefont {Rechtsman}}, \bibinfo {author} {\bibfnamefont {D.}~\bibnamefont
  {Schuster}}, \bibinfo {author} {\bibfnamefont {J.}~\bibnamefont {Simon}},
  \bibinfo {author} {\bibfnamefont {O.}~\bibnamefont {Zilberberg}},\ and\
  \bibinfo {author} {\bibfnamefont {I.}~\bibnamefont {Carusotto}},\ }\bibfield
  {title} {\bibinfo {title} {Topological photonics},\ }\href@noop {} {\bibfield
   {journal} {\bibinfo  {journal} {Rev. Mod. Phys.}\ }\textbf {\bibinfo
  {volume} {91}},\ \bibinfo {pages} {015006} (\bibinfo {year}
  {2019})}\BibitemShut {NoStop}%
\bibitem [{\citenamefont {Schwartz}\ \emph {et~al.}(2007)\citenamefont
  {Schwartz}, \citenamefont {Bartal}, \citenamefont {Fishman},\ and\
  \citenamefont {Segev}}]{Schwartz2007}%
  \BibitemOpen
  \bibfield  {author} {\bibinfo {author} {\bibfnamefont {T.}~\bibnamefont
  {Schwartz}}, \bibinfo {author} {\bibfnamefont {G.}~\bibnamefont {Bartal}},
  \bibinfo {author} {\bibfnamefont {S.}~\bibnamefont {Fishman}},\ and\ \bibinfo
  {author} {\bibfnamefont {M.}~\bibnamefont {Segev}},\ }\bibfield  {title}
  {\bibinfo {title} {Transport and anderson localization in disordered
  two-dimensional photonic lattices},\ }\href@noop {} {\bibfield  {journal}
  {\bibinfo  {journal} {Nature}\ }\textbf {\bibinfo {volume} {446}},\ \bibinfo
  {pages} {52} (\bibinfo {year} {2007})}\BibitemShut {NoStop}%
\bibitem [{\citenamefont {Rechtsman}\ \emph {et~al.}(2013)\citenamefont
  {Rechtsman}, \citenamefont {Zeuner}, \citenamefont {Plotnik}, \citenamefont
  {Lumer}, \citenamefont {Podolsky}, \citenamefont {Dreisow}, \citenamefont
  {Nolte}, \citenamefont {Segev},\ and\ \citenamefont
  {Szameit}}]{Rechtsman2013}%
  \BibitemOpen
  \bibfield  {author} {\bibinfo {author} {\bibfnamefont {M.~C.}\ \bibnamefont
  {Rechtsman}}, \bibinfo {author} {\bibfnamefont {J.~M.}\ \bibnamefont
  {Zeuner}}, \bibinfo {author} {\bibfnamefont {Y.}~\bibnamefont {Plotnik}},
  \bibinfo {author} {\bibfnamefont {Y.}~\bibnamefont {Lumer}}, \bibinfo
  {author} {\bibfnamefont {D.}~\bibnamefont {Podolsky}}, \bibinfo {author}
  {\bibfnamefont {F.}~\bibnamefont {Dreisow}}, \bibinfo {author} {\bibfnamefont
  {S.}~\bibnamefont {Nolte}}, \bibinfo {author} {\bibfnamefont
  {M.}~\bibnamefont {Segev}},\ and\ \bibinfo {author} {\bibfnamefont
  {A.}~\bibnamefont {Szameit}},\ }\bibfield  {title} {\bibinfo {title}
  {Photonic floquet topological insulators},\ }\href@noop {} {\bibfield
  {journal} {\bibinfo  {journal} {Nature}\ }\textbf {\bibinfo {volume} {496}},\
  \bibinfo {pages} {196} (\bibinfo {year} {2013})}\BibitemShut {NoStop}%
\bibitem [{\citenamefont {Savona}\ \emph {et~al.}(1995)\citenamefont {Savona},
  \citenamefont {Andreani}, \citenamefont {Schwendimann},\ and\ \citenamefont
  {Quattropani}}]{savona1995quantum}%
  \BibitemOpen
  \bibfield  {author} {\bibinfo {author} {\bibfnamefont {V.}~\bibnamefont
  {Savona}}, \bibinfo {author} {\bibfnamefont {L.}~\bibnamefont {Andreani}},
  \bibinfo {author} {\bibfnamefont {P.}~\bibnamefont {Schwendimann}},\ and\
  \bibinfo {author} {\bibfnamefont {A.}~\bibnamefont {Quattropani}},\
  }\bibfield  {title} {\bibinfo {title} {Quantum well excitons in semiconductor
  microcavities: Unified treatment of weak and strong coupling regimes},\
  }\href@noop {} {\bibfield  {journal} {\bibinfo  {journal} {Solid State
  Commun.}\ }\textbf {\bibinfo {volume} {93}},\ \bibinfo {pages} {733}
  (\bibinfo {year} {1995})}\BibitemShut {NoStop}%
\bibitem [{\citenamefont {Hopfield}(1958)}]{hopfield1958theory}%
  \BibitemOpen
  \bibfield  {author} {\bibinfo {author} {\bibfnamefont {J.}~\bibnamefont
  {Hopfield}},\ }\bibfield  {title} {\bibinfo {title} {Theory of the
  contribution of excitons to the complex dielectric constant of crystals},\
  }\href@noop {} {\bibfield  {journal} {\bibinfo  {journal} {Phys. Rev.}\
  }\textbf {\bibinfo {volume} {112}},\ \bibinfo {pages} {1555} (\bibinfo {year}
  {1958})}\BibitemShut {NoStop}%
\end{thebibliography}

%

\end{document}